\documentclass[lettersize,journal]{IEEEtran}
\usepackage{amsmath,amsfonts}
\usepackage{algorithmic}
\usepackage{algorithm}
\usepackage{amssymb}
\usepackage{array}
\usepackage{bm}
\usepackage{color}
\usepackage{makecell}
\usepackage{multirow}
\usepackage[caption=false,font=normalsize,labelfont=sf,textfont=sf]{subfig}
\usepackage{textcomp}
\usepackage{stfloats}
\usepackage{url}
\usepackage{verbatim}
\usepackage{graphicx}
\usepackage{cite}
\begin{document}

\title{\fontsize{20pt}{20pt}\selectfont Integrated Semantic Sensing and Communications: Target Attribute Inference and Beamforming Design}

\author{Haobo Zhang,~\IEEEmembership{Member,~IEEE,}
        Shuhao~Zeng,~\IEEEmembership{Member,~IEEE,}
        Boya~Di,~\IEEEmembership{Senior Member,~IEEE,}
        Hongliang~Zhang,~\IEEEmembership{Member,~IEEE,}
        Harald~Haas,~\IEEEmembership{Fellow,~IEEE}
        and~Lingyang~Song,~\IEEEmembership{Fellow,~IEEE}
        \vspace{-5mm}
\thanks{H. Zhang is with School of Electronic and Computer Engineering, Peking University Shenzhen Graduate School, Shenzhen 518055, China, and also with the Department of Engineering, Electrical Engineering Division, Cambridge University, Cambridge CB3 0FA, UK (email: haobo.zh97@gmail.com).}%
\thanks{S. Zeng is with School of Electronic and Computer Engineering, Peking University Shenzhen Graduate School, Shenzhen, China, and also with Department of Electrical and Computer Engineering, Princeton University, NJ, USA (email: shuhao.zeng96@gmail.com).}
\thanks{B. Di and H. Zhang are with State Key Laboratory of Photonics and Communications, School of Electronics, Peking University, Beijing 100871, China (e-mail: \{diboya,hongliang.zhang\}@pku.edu.cn).}%
\thanks{H. Haas is with the Department of Engineering, Electrical Engineering Division, Cambridge University, Cambridge CB3 0FA, UK (e-mail:huh21@cam.ac.uk).}
\thanks{L. Song is with School of Electronic and Computer Engineering, Peking University Shenzhen Graduate School, Shenzhen 518055, China, and also with State Key Laboratory of Photonics and Communications, School of Electronics, Peking University, Beijing 100871, China (email: lingyang.song@pku.edu.cn).}%
}

\markboth{IEEE TRANSACTIONS ON WIRELESS COMMUNICATIONS,~Vol.~14, No.~8, August~2026}%
{Shell \MakeLowercase{\textit{et al.}}: A Sample Article Using IEEEtran.cls for IEEE Journals}

\maketitle

\begin{abstract}
Integrated sensing and communication (ISAC) is expected to become a key enabling paradigm for future 6G networks, supporting emerging applications that require a comprehensive understanding of sensed targets.
As a step toward this goal, we consider an ISAC system with semantic sensing capability that determines what the sensed target is, where it is located, and how large it is.
However, the accurate retrieval of these attributes, including target class, location, and size, is challenging due to the highly nonlinear relationships between the received signals and different attributes. In addition, new beamforming strategies are also required to balance multi-user communication and semantic sensing performance. To address the first challenge, we develop a physics-guided inference model that leverages signal propagation priors and a unified Transformer backbone to effectively characterize the highly nonlinear relationships between received ISAC signals and multiple attributes.
As for the second challenge, we develop a multi-stage training framework for the optimization of beamformers and model parameters, aiming to improve semantic sensing performance while guaranteeing multi-user communication requirements. Simulation results show that under the same multi-user communication rate requirements, the proposed scheme can provide more than 12\% performance gains across classification, localization, and size estimation tasks compared with existing methods and achieve 91.21\% classification accuracy using only two OFDM frames, demonstrating the feasibility of accurate semantic sensing with low measurement overhead.
\end{abstract}


\vspace{-2mm}
\section{Introduction}

\IEEEPARstart{I}{tegrated} sensing and communication (ISAC) has emerged as a key enabling paradigm for future 6G networks, aiming to unify wireless sensing and communication within a single hardware platform, shared spectrum, and joint signal processing framework. By exploiting the dual functionality of radio signals, ISAC enables simultaneous environment perception and data transmission, thereby significantly improving spectrum efficiency and reducing hardware cost~\cite{zhang2026integrated,zhang2026holographic}.

In the literature, various ISAC schemes have been developed to acquire different attributes of sensed targets. In~\cite{zhang2025target}, the authors investigated the estimation of target locations, and a two-stage ISAC scheme was proposed to reuse communication signals for localization.  
Moreover, existing works also explored the effectiveness of target classification using ISAC systems. For example, the authors in~\cite{guo2025multiges} considered a human-computer interaction application, where the human gesture recognition is realized using ISAC signals, and the authors in~\cite{yin2025deep} studied the classification of different UAV types using orthogonal frequency-division multiplexing~(OFDM) signals.

\textcolor{black}{However, emerging applications often require joint knowledge of multiple target semantics, which refer to diverse high-level meaningful attributes of a target such as its class, location, and size~\cite{rosinol2020kimera, he2026task}. ISAC designs that provide only a single attribute may thus result in incorrect decisions or even task failure. For example, an ISAC-enabled smart home network~\cite{shakoor2025simultaneous} needs both target class and location to monitor human presence within a designated zone. A location estimate alone cannot distinguish people from other targets, whereas the target class information alone cannot determine whether a person is within that zone. This limitation becomes more pronounced in factory intrusion monitoring~\cite{tosi2024feasibility}, where ISAC networks must consider target size in addition to class and location. Neglecting target size may leave an intrusion undetected when part of the target enters a protected area while its center remains outside. Moreover, in sensing-assisted communication, incomplete knowledge of potential obstacles, including their classes, locations, and sizes, may compromise predictions of blockage occurrence and duration, hindering timely beam switching and increasing the risk of communication outages~\cite{wu2023machine}.}

Motivated by the above information requirements, we investigate the integration of semantic sensing capability~\cite{gao2026generative} into ISAC systems in this paper, where semantic sensing aims to directly infer multiple target attributes from signals by jointly answering what the target is, where it is located, and how large it is. Specifically, we consider an OFDM-based ISAC system, where a base station (BS) transmits ISAC signals to simultaneously support communication and semantic sensing. The signals reflected by the sensed target are collected by a passive radar (PR) and leveraged to infer target attributes, including target class, location, and size.

Enabling semantic sensing in ISAC systems is, however, highly challenging. First, the accurate retrieval of target attributes from received ISAC signals is hindered by the highly nonlinear relationships between received signals and different target attributes. In addition, new beamforming strategies are required for integrated semantic sensing and communication, since they need to balance communication rate with semantic sensing performance that simultaneously accounts for classification, localization, and size estimation accuracy.

To address the first challenge, a physics-guided target attribute inference model based on a unified Transformer backbone is developed to effectively capture the complex relationships between received signals and different target attributes. As for the second challenge, a multi-stage training framework is proposed, where the beamformers and model parameters are optimized to improve the semantic sensing performance under communication constraints. Overall, the main contributions of this paper are summarized as follows:

\begin{itemize}
    \item We propose an integrated semantic sensing and communication framework that simultaneously supports semantic sensing and multi-user communication. As a low-cost and power-efficient implementation, we employ a reconfigurable intelligent surface~(RIS)-assisted BS to generate customized ISAC beams, while a passive radar (PR) collects target-reflected signals for the joint inference of target classes, locations, and dimensions.
    \item We develop a physics-guided inference model for target attribute extraction from received ISAC signals. Motivated by the observation that target location predominantly determines the large-scale responses across subcarriers, while the detailed target structures introduce small-scale signal variations, the proposed framework designs dedicated tokens to decouple these effects and generate complementary tokens for inference. These tokens are subsequently processed by a unified Transformer backbone to directly infer target class, location, and size.
    \item We propose a multi-stage training framework for the joint beamformer design and inference model fine-tuning. Specifically, the inference model is first pre-trained, while the digital and RIS beamformers are also initialized. Then, the entire system is jointly fine-tuned in an end-to-end manner to improve semantic sensing performance while satisfying multi-user communication constraints.
    \item Simulation results verify the effectiveness of the proposed scheme. Specifically, it achieves 91.21\% classification accuracy using only two OFDM frames and provides more than 12\% performance gains for classification, localization, and size estimation under the same communication rate constraint compared with existing schemes, demonstrating the feasibility of accurate semantic sensing with low measurement overhead.
\end{itemize}

The rest of the paper is organized as follows. The related works are first elaborated on in Sec.~\ref{s_rw}. In Sec.~\ref{s_sm}, the mathematical models for integrated semantic sensing and communication are provided. The design of target attribute inference network is introduced in Sec.~\ref{s_rasi}, and the multi-stage training framework is described in Sec.~\ref{s_mstf}. The numerical results are provided in Sec.~\ref{s_nr}, and conclusions are drawn in Sec.~\ref{s_c}.

\vspace{-2mm}
\section{Related Works}
\label{s_rw}

In this section, we first review the related works on wireless sensing in ISAC systems. Next, another relevant topic, i.e., integrated sensing and semantic communications, is introduced and compared with the proposed scheme.

\vspace{-3mm}
\subsection{Sensing in ISAC Systems}

Sensing capability is essential in ISAC systems to support environment awareness and enhance communication performance. In conventional ISAC systems, multiple-input multiple-output~(MIMO) antenna arrays are widely adopted for sensing since they enable the separation of target echo signals arriving from different directions. The estimation of various target attributes using antenna array have been investigated, including location~\cite{zhang2025joint}, reflection coefficient~\cite{zhang2022parameter}, and class~\cite{yin2025deep}. 

Recently, RISs have been widely adopted in ISAC systems as a low cost and power efficient implementation of massive MIMO~\cite{zhong2023joint,zeng2026holograhpic}. They can provide large apertures using passive reflecting elements, thereby significantly reducing hardware cost and power consumption compared with conventional active antenna arrays. For example, the authors in~\cite{zhang2022metaradar} proposed to employ RIS to customize the wireless channels between the radar and targets, and radar waveforms and RIS beamformers were jointly optimized to improve the multi-target detection performance. The minimization of CRB given the communication users' signal-to-interference-plus-noise ratio~(SINR) constraints in an active RIS-aided system was investigated in~\cite{zhu2024cramer}. More recently, RIS has been employed for semantic segmentation\footnote{The semantic segmentation task considered in~\cite{hu2022metasketch} is different from the object-level semantic sensing task considered in this paper. Specifically, the authors in~\cite{hu2022metasketch} employed RIS for environmental imaging via compressive sensing and further assigned semantic labels to individual image pixels, and the divergence between the estimated semantic labels and the ground truth was evaluated. In contrast, this paper focuses on directly estimating the object-level location, size, and class attribute rather than recovering the whole image. In addition, the system in~\cite{hu2022metasketch} is optimized for sensing and thus cannot guarantee the communication performance.}~\cite{hu2022metasketch}. These studies demonstrate the potential of RIS in enabling power- and cost-efficient wireless sensing~\cite{hu2024cross}.

Despite these advances, existing wireless sensing systems still exhibit limitations for semantic sensing. Specifically, existing methods are mainly designed and optimized for specific sensing tasks, such as localization or classification. Thus, directly applying these methods to semantic sensing does not guarantee the performance of jointly inferring multiple target attributes, and the optimization of RIS and digital beamformers for balancing communication and semantic sensing performance remains largely unexplored. Although~\cite{zhao2026extended} considered joint classification and localization, the sensing architecture relies on large-scale active antenna arrays instead of RISs, and the target size estimation was not incorporated. Motivated by these limitations, this paper investigates integrated semantic sensing and communication aided by RIS, which aims to jointly infer target class, location, and size under multi-user communication constraints.

\color{black}

\vspace{-3mm}
\subsection{Semantic Communication}

Semantic communication has been proposed as a paradigm that departs from conventional bit-level transmission by focusing on task-relevant or meaning-level information delivery~\cite{zhang2026feature}. It mainly aims to improve communication efficiency, robustness, or security by transmitting semantic representations instead of raw data. Some works also investigate how to integrate semantic communication into ISAC systems. For example, the authors in~\cite{yang2025integrated} considered an ISAC system with a roadside unit, multiple unintended vehicles and intended vehicles, where the roadside unit performed semantic communication with all the intended vehicles and tracked all the vehicles. The ISAC waveforms were optimized to maximize the semantic secrecy rate and minimize the Cram\'er-Rao bound of sensing paramters such as angle and path loss.

Nevertheless, existing studies on integrated semantic communication and sensing differ fundamentally from the considered framework. Existing works mainly integrate semantic feature extraction with communication transmission, where semantic representations are extracted from source data and delivered to users through semantic communication techniques. In contrast, this paper explores another paradigm by integrating semantic-aware inference with the sensing process itself, where target attributes, including target class, location, and size, are directly inferred from received signals rather than being extracted for transmission purposes. Furthermore, this paper investigates the joint optimization of beamforming strategies tailored for semantic-aware inference while guaranteeing multi-user communication rates.

\vspace{-1mm}
\section{System Model}
\label{s_sm}

In this section, we first describe the integrated semantic sensing and communication scenario, and then introduce the signal model and protocol, respectively.

\vspace{-3mm}
\subsection{Scenario Description}

\begin{figure}[!t]
  \centering
  \includegraphics[width=2.8in]{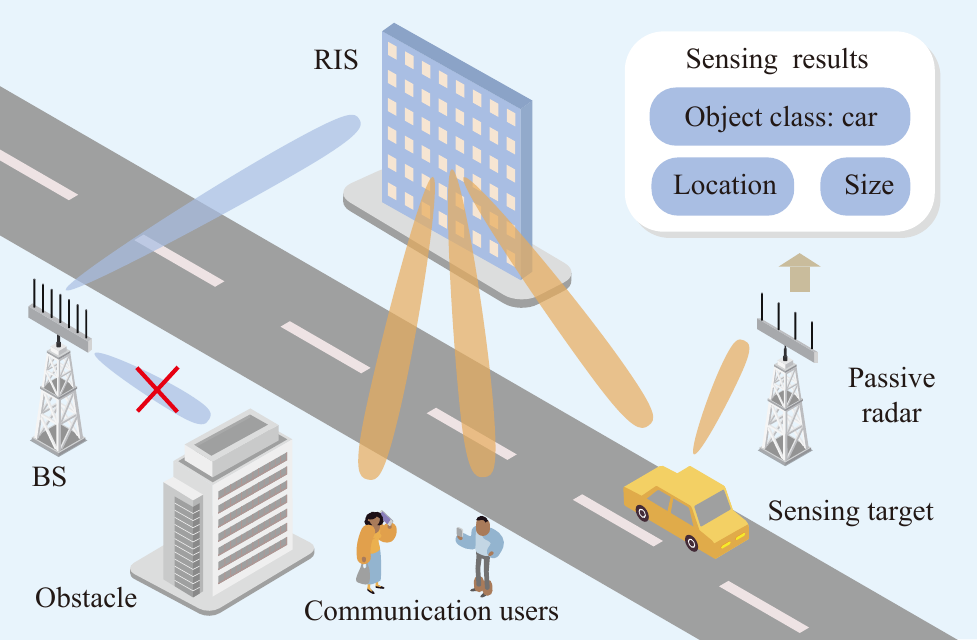}
  \caption{Illustration of integrated semantic sensing and communication scenario.}
  \label{f_scenario}
  \vspace{-4mm}
\end{figure}

As shown in Fig.~\ref{f_scenario}, we consider an integrated semantic sensing and communication system, which is composed of a BS with $M$ antennas, a passive radar~(PR) with $K$ antennas, a RIS with $N$ elements, $U$ single-antenna users, and a sensing target.
The line-of-sight (LoS) paths between the BS and the users/target are blocked due to the existence of obstacles, and the BS performs the communication and semantic sensing tasks via the cooperation with RIS and PR~\cite{bazzi2025low, zhang2021metalocalization}.

Specifically, to perform semantic sensing and communication simultaneously, the BS first transmits the ISAC signals via its antenna array towards the RIS, which reflects the signals to the users and the target. The users then decode the received signals to recover the communication information. At the same time, the PR records the signals scattered from the target and estimates the attributes of the target.

\subsection{Signal Model}

An orthogonal frequency division multiplexing~(OFDM) waveform with $S$ subcarriers is adopted in the integrated semantic sensing and communication system. Suppose a transmission frame composed of $L$ successive OFDM symbols for ISAC, and let $\bm{D}_s \in \mathbb{C}^{U \times L}$ and $\bm{W}_s \in \mathbb{C}^{M \times L}$ denote the communication data symbol and sensing waveform for the $s$-th subcarrier. The transmitted signal at the $s$-th subcarrier can then be given by~\cite{xiao2025sparsity, zhang2022holographic}
\begin{equation}
    \bm{X}_s = \bm{B}^{\text{comm}}_s \bm{D}_s + \bm{B}^{\text{sen}}_s \bm{W}_s,
\end{equation}
where $\bm{B}^{\text{comm}}_s \in \mathbb{C}^{M \times U}$ and $\bm{B}^{\text{sen}}_s \in \mathbb{C}^{M \times M}$ are the digital beamformers for communication and sensing, respectively. In addition, we define the overall digital beamformer as $\bm{B}_s = (\bm{B}^{\text{comm}}_s, \bm{B}^{\text{sen}}_s) \in \mathbb{C}^{M \times (U+M)}$.
Since there is no direct link between the BS and the users, the received signal of all the $U$ users at the $s$-th subcarrier can be expressed as
\begin{equation}
  \bm{Y}_s = \bm{H}^{\text{user}}_s \bm{\Psi} \bm{H}^{\text{BS}}_s \bm{X}_s + \bm{N}^{\text{user}}_s,
\end{equation}
where $\bm{H}^{\text{user}}_m \in \mathbb{C}^{U \times N}$ is the channel between the RIS and the users for the $s$-th subcarrier, $\bm{H}^{\text{BS}}_m \in \mathbb{C}^{N \times M}$ is the channel between the BS and the RIS for the $s$-th subcarrier, $\bm{N}^{\text{user}}_m \in \mathbb{C}^{U \times L}$ is the noise at the user antennas for the $s$-th subcarrier, and $\bm{\Psi} \in \mathbb{C}^{N \times N}$ is the reflection matrix of the RIS, which can be given by
\begin{equation}
  \bm{\Psi} = \text{diag} \left( \beta e^{j\psi_1}, \beta e^{j\psi_2}, \cdots, \beta e^{j\psi_N} \right).
\end{equation}
Here, $\beta$ is the amplitude of the reflection coefficient, and $\psi_n$ is the phase shift of the $n$-th RIS element. We assume that the RIS operating bandwidth is sufficiently large, such that the phase shifts of the RIS elements can be regarded as frequency-independent across all subcarriers~\cite{sheng2025single, zeng2024ris}. The beamformer vector of the RIS is denoted by $\bm{\psi} = (\psi_1, \psi_2, \cdots, \psi_N)^{\mathsf T}$. 

Given the received signal, the signal-to-interference-plus-noise ratio~(SINR) of the $u$-th user can be given by
\begin{align}
    \gamma_{s, u}
    = \dfrac{|(\bm{h}^{\text{user}}_{s, u})^{\mathsf H} \bm{\Psi} \bm{H}^{\text{BS}}_s \bm{b}^{\text{comm}}_{s, u}|^2}{I_{s, u}+(\sigma^{\text{user}}_{s, u})^2},\label{d_gamma}
\end{align}
where $\bm{h}^{\text{user}}_{s, u}$ is the $u$-th row of $\bm{H}^{\text{user}}_s$ and represents the channel between the RIS and the $u$-th user, $\bm{b}^{\text{comm}}_{s, u}$ is the $u$-th column of $\bm{B}^{\text{comm}}_s$ and represents the beamformer for the $u$-th user, $\sigma^{\text{user}}_{s, u}$ is the standard deviation of noise, and $I_{s, u}$ is the interference for the $u$-th user at the $s$-th subcarrier, which is given by
\begin{align}
    I_{s, u} =~& \sum_{u'\ne u}\left|(\bm{h}^{\text{user}}_{s, u})^{\mathsf H} \bm{\Psi} \bm{H}^{\text{BS}}_s \bm{b}^{\text{comm}}_{s, u'}\right|^2 \notag\\
    &+ \left\|(\bm{h}^{\text{user}}_{s, u})^{\mathsf H} \bm{\Psi} \bm{H}^{\text{BS}}_s \bm{B}^{\text{sen}}_s\right\|^2_2.
\end{align}
Consequently, the average communication rate of the $u$-user is given by
\begin{equation}
    R_u = \dfrac{1}{S}\sum^S_{s=1} \log_2(1+\gamma_{s, u}).\label{d_r_k}
\end{equation}

In addition, the target in the sensing area also reflects the signal from the RIS to the PR. Here, we discrete the whole sensing area into voxels. The size of each voxel is $\xi_x \times \xi_y \times \xi_z$, and the scattering coefficient of the $i$-th voexl is $v_i$. Thus, the signal received by the PR for the $s$-th subcarrier can be expressed as
\begin{equation}
  \bm{Z}_s = \bm{H}^{\text{PR}}_s \bm{\Psi} \bm{H}^{\text{BS}}_s \bm{X}_s + \bm{N}^{\text{PR}}_s, \label{d_Z_m}
\end{equation} 
where $\bm{H}^{\text{PR}}_s \in \mathbb{C}^{K \times N}$ is the channel from RIS to the PR, and $\bm{N}^{\text{PR}}_s \in \mathbb{C}^{K \times N}$ is the noise at the PR. The RIS-PR channel is given by
\begin{equation}
    \bm{H}^{\text{PR}}_s = \bm{H}^{\text{V-PR}}_s \text{diag}(\bm{v}) \bm{H}^{\text{RIS-V}}_s.
\end{equation}
Here, $\bm{H}^{RIS-V}_s \in \mathbb{C}^{V \times N}$ is the channel between the RIS and the voxels, $V$ is the number of voxels, $\bm{v} = (v_1, v_2, \cdots, v_V)^{\mathsf H}$ is the scattering vector, $\bm{H}^{V-PR}_s \in \mathbb{C}^{K \times V}$ is the channel between the voxels and PR.

\subsection{Integrated Semantic Sensing and Communication Protocol}

\begin{figure}[!t]
  \centering
  \includegraphics[width=3.2in]{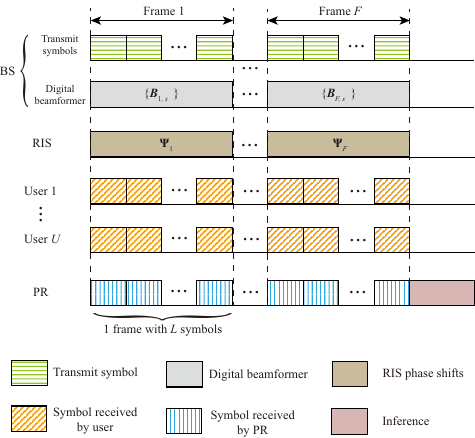}
  \caption{Integrated semantic sensing and communication protocol.}
  \label{f_protocol}
  \vspace{-4mm}
\end{figure}

The proposed integrated semantic sensing and communication protocol is illustrated in Fig.~\ref{f_protocol}. Without loss of generality, we assume that the protocol consists of $F \ge 1$ frames. 
In each frame, $L$ symbols are transmitted for joint semantic sensing and downlink communication. After $F$ frames, the target attributes are estimated from the received signals across all the frames using the proposed learning model. Details are elaborated in the following.

\subsubsection{ISAC Transmission}

In the $f$-th frame, the digital beamformers for semantic sensing and communication are denoted by $\{\bm{B}_{f, s}\}^S_{s=1}$, where $\bm{B}_{f, s}$ is the digital beamformer at the $s$-th subcarrier in the $f$-th frame. The digital beamformers are employed to generated the transmitted OFDM symbols $\{\bm{X}_{f, s}\}^S_{s=1}$. The signals from the BS arrive at the users and PR via the reflection of RIS, whose beamformer is $\bm{\psi}_f$. Note that the digital and RIS beamformers for semantic sensing and communication vary across different frames, and let $\bm{B}\in \mathbb{C}^{F \times S \times M  \times (M+U)}$ and $\bm{\psi}\in \mathbb{C}^{F \times N}$ denote all the digital and RIS beamformers, respectively. To improve the performance of communication and semantic sensing, the RIS beamformer $\bm{\Psi}$ and the digital beamformers $\bm{B}$ need to be optimized, which are discussed in detail in Sec.~\ref{s_mstf}.

\subsubsection{Signal Reception}

In the $f$-th frame, the signals received by all the users is given by $\{\bm{Y}_{f, s}\}^S_{s=1}$, where $\bm{Y}_{f, s}$ is the received symbols at the $s$-th subcarrier in the $f$-th frame. The communication data is retrieved from $\{\bm{Y}_{f, s}\}^S_{s=1}$ to support downlink transmission, and the average communication rate of the $u$-th user can be calculated based on~(\ref{d_r_k}). As for the PR, the received signal is $\{\bm{Z}_{f, s}\}^S_{s=1}$, where $\bm{Z}_{f, s}$ is the signals received by PR at the $s$-th subcarrier in the $f$-th frame. For the PR, the interaction between the wireless signals and the target implicitly embeds target attributes, including class, location, and size, into the received waveforms.

\subsubsection{Target Attribute Inference}

In this step, target attributes are directly inferred based on the received signal. 

In this paper, we consider three types of target attributes, i.e., $\bm{o} = (c, \bm{r}, \bm{s})$, where $c$, $\bm{r}$, and $\bm{s}$ are the class, location, and size of the target. Here, the considered extended target corresponds to multiple voxels with non-zero scattering coefficients. Instead of reconstructing the complete shape of the target, an inference model that directly outputs the attributes $\bm{o}$ is designed. The details of the inference model are presented in Sec.~\ref{s_rasi}.

\section{Physics-Guided Target Attribute Inference}
\label{s_rasi}

In this section, we propose a physics-guided inference algorithm that estimates the target attributes from the received OFDM signals, transmit symbols, digital beamformers, and RIS beamformers. The overview of the proposed algorithm is first introduced in Sec.~\ref{ss_ao}, and the modules that constitute the algorithm are described in \ref{ss_fem} and \ref{ss_sim}, respectively.

\subsection{Algorithm Overview}
\label{ss_ao}

The aim of the proposed inference algorithm is to estimate the attributes $\bm{o}$ of the extended target, including its class $c$, location $\bm{r}$, and size $\bm{s}$. Let $\mathcal{C}$ denote the set containing all the target classes, and we have $c \in \mathcal{C}$.
Besides, suppose the whole sensing region is discretized into multiple voxels with the same size, and the indexes of voxels occupied by the target is denoted by $\mathcal{V}$. Since all occupied voxels have identical sizes, we define the target location as the geometric center of the occupied voxels. The size of the target is $\bm{s} = (s_x, s_y, s_z)$, where $s_x$, $s_y$, and $s_z$ are the lengths of the target along the $x$, $y$, and $z$ axes, respectively. Here, the length $s_x$ can be given by
\begin{align}
    s_x =~& \max_{i, j} |r_{x, i} - r_{x, j}|, \forall i, j \in \mathcal{V},
\end{align}
where $r_{x, i}$ and $r_{x, j}$ are the $x$- and $y$-coordinates of the center of the $i$-th voxel, respectively, and the lengths along other axes can be calculated in similar ways. 

\begin{figure*}[!t]
  \centering
  \includegraphics[width=6in]{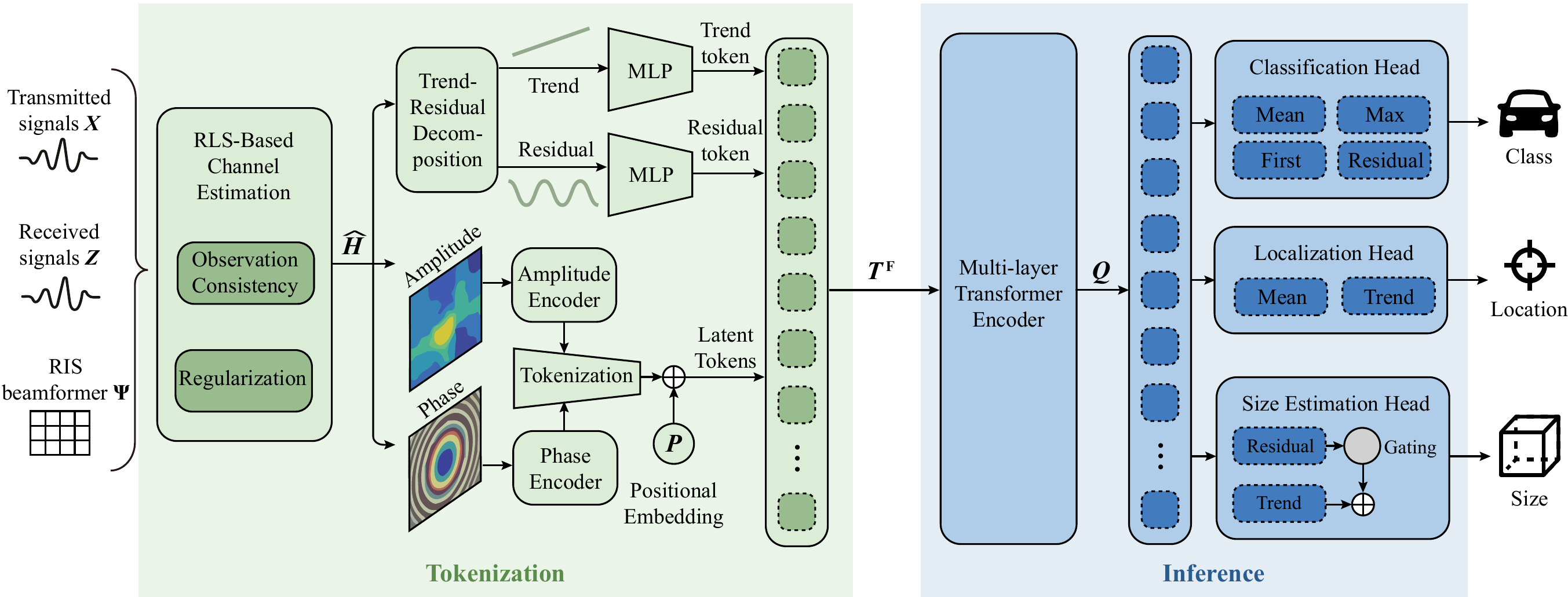}
  \caption{The architecture of the proposed inference algorithm.}
  \label{f_algorithm}
  \vspace{-4mm}
\end{figure*}

To achieve the aforementioned aim, we propose a deep learning-based network that directly estimates the target attributes from the received signals. The proposed model consists of two cascaded modules, i.e., the physics-guided tokenization, and transformer-based inference modules. The first module is employed to convert wireless signals to tokens, which are then input into the inference module to obtain the task-oriented attributes of the target.

\subsection{Physics-Guided Tokenization Module}
\label{ss_fem}

The objective of this module is to generate token representations $\bm{T}^{\rm F}$ from the transmitted signals $\bm{X} \in \mathbb{C}^{F \times S \times M \times L}$, the received signals $\bm{Z} \in \mathbb{C}^{F \times S \times K \times L}$, and the RIS reflection matrix $\bm{\Psi} \in \mathbb{C}^{F \times N \times N}$\footnote{For notational simplicity, the batch dimension is omitted in this section.}. The proposed physics-guided tokenization module exploits physical priors inherent in wireless signal propagation by explicitly characterizing both large-scale responses associated with target location and fine-grained variations induced by local scattering structures. Specifically, the module first estimate the RIS–PR channel $\widehat{\bm{H}} \in\mathbb{C}^{S \times K \times N}$ using regularized least squares~(RLS)-based method with carefully designed regularization terms. A trend-residual decomposition mechanism is then introduced to disentangle large-scale responses and fine-grained signal variations, generating structural tokens that encode complementary physical characteristics of the sensing target. In addition, latent tokens are extracted from the amplitude and phase components of the estimated channel to provide supplementary information for target attribute inference. Finally, the structural tokens are concatenated with the latent feature tokens to form the complete token sequence $\bm{T}^{\rm F}$.

\subsubsection{RLS-Based Channel Estimation}

According to the signal model~(\ref{d_Z_m}), the received signal at the $s$-th subcarrier in the $f$-th frame can be written as
\begin{equation}
\bm{Z}^{\mathsf H}_{f, s} = \bm{A}_{f, s} \left(\bm{H}^{\text{PR}}_s\right)^{\mathsf H} + \left(\bm{N}^{\text{PR}}_{f, s}\right)^{\mathsf H},\label{d_Z_f_s}
\end{equation}
where $\bm{H}^{\text{PR}}_s$ is the RIS–receiver channel to be estimated, and $\bm{A}_{f, s}$ is the measurement matrix for the $f$-th frame at the $s$-th subcarrier, given by
\begin{equation}
    \bm{A}_{f, s} = \left(\bm{\Psi}_f \bm{H}^{\text{BS}}_s \bm{X}_{f, s}\right)^{\mathsf H}
\end{equation}
Here, reflection matrix $\bm{\Psi}_f$ and transmitted signal $\bm{X}_{f, s}$ are known, and the BS-RIS channel $\bm{H}^{\text{BS}}_s$ can be obtained using existing channel estimation method~\cite{fredj2024channel}.
Since (\ref{d_Z_f_s}) can be viewed as a linear inverse problem with respect to $\bm{H}^{\text{PR}}_s$, the channel can be estimated by solving a least squares~(LS) problem. To improve robustness under noise and limited observations, we formulate a regularized LS problem as
\begin{align}
\widetilde{\bm{H}}_s
=~& \arg\min_{\bm{H}}
\sum^F_{f=1}||\mathbf{A}_{f,s}\bm{H}^{\mathsf H}-\bm{Z}^{\mathsf H}_{f,s}||^2_F
+ \lambda_r ||\bm{H}||^2_F
\notag\\
&+ \lambda_s||\bm{L}\bm{H}^{\mathsf H}||^2_F,\label{d_H_PR}
\end{align}
where $\widetilde{\bm{H}}_s \in \mathbb{C}^{K \times N}$ is the estimated RIS-PR channel over the $s$-th subcarrier, $\lambda_r$ and $\lambda_s$ are two coefficients for the regularization terms, respectively, and $\bm{L}$ denotes the Laplacian matrix defined on the two-dimensional RIS grid.

In~(\ref{d_H_PR}), the first term $||\mathbf{A}_{f,s}\bm{H}^{\mathsf H}-\bm{Z}^{\mathsf H}_{f,s}||^2_F$ enforces consistency between the observations and the estimated signals, while the second and third terms correspond to the ridge and spatial regularization terms, respectively. Here, the second term $\lambda_r||\bm{H}||_F^2$ is introduced to stabilize the inversion~\cite{xiao2026channel}. Specifically, when the measurements collected across the $F$ frames provide
insufficient independent information, the unregularized least-squares
estimate may be highly sensitive to noise, such that small perturbations in the received signals may lead to large errors in the estimated channel. The ridge term alleviates this issue by penalizing excessively large channel coefficients and improving the robustness of the channel estimation.

In addition, the third term $\lambda_s||\bm{L}\bm{H}^{\mathsf H}||_F^2$ further exploits the spatial structure of the RIS aperture. It can suppress undesirable components with high spatial frequencies in the estimated channel, which are also likely to arise from noise amplification and the ill-conditioning of the inverse problem. To define $\bm{L}$ in (\ref{d_H_PR}), we first model the RIS as a two-dimensional grid graph, where each RIS element corresponds to a node and two nodes are connected if the corresponding elements are adjacent on the aperture. Let $\bm{L}' \in \mathbb{R}^{N\times N}$ denote the adjacency matrix, where $L'_{i,j}=1$ if elements $i$ and $j$ are neighbors, and otherwise $L'_{ij}=0$. Besides, let $\bm{D}$ be the diagonal degree matrix with $D_{i,i}=\sum_j L'_{i j}$. Then, the Laplacian matrix is given by~\cite{egilmez2017graph}
\begin{equation}
\bm{L}=\bm{D}-\bm{L}'.
\end{equation}
The term $||\bm{L}\bm{H}^{\mathsf H}||_F^2$ measures local differences between neighboring elements. Thus, it can penalize high-frequency variations of the estimated channel over the RIS grid and encourages spatially consistent solutions.

The optimization problem in (\ref{d_H_PR}) is an unconstrained quadratic problem with respect to $\bm{H}$ and thus admits a closed-form solution. By setting the derivative of the objective function to zero, we have
\begin{equation}
\begin{aligned}
\left(
\sum_{f=1}^{F}\bm A_{f,s}^H\bm A_{f,s}
+\lambda_r\bm I_N
+\lambda_s\bm L^H\bm L
\right)
\widehat{\bm H}_s^H
=
\sum_{f=1}^{F}\bm A_{f,s}^H\bm Z_{f,s}^H.
\end{aligned}
\end{equation}
Accordingly, the solution can be expressed as
\begin{equation}
\begin{aligned}
\widehat{\mathbf H}_s^H
={}&
\left(
\sum_{f=1}^{F}\mathbf A_{f,s}^H\mathbf A_{f,s}
+\lambda_r\mathbf I_N
+\lambda_s\mathbf L^H\mathbf L
\right)^{-1} \\
&\times
\left(
\sum_{f=1}^{F}\mathbf A_{f,s}^H\mathbf Z_{f,s}^H
\right).
\end{aligned}
\end{equation}

\subsubsection{Trend and Residual Token Generation}

To encode physically meaningful channel structure, the channel is processed by a trend-residual decomposition operation. Let $\bm{\varphi}_{k, n} \in \mathbb{R}^{S}$ denote the phase of the channel across all subcarriers. The module first performs phase unwrapping along the frequency axis and then fits a linear trend
\begin{equation}
\bm{\varphi}^{\rm tr}_{k, n} = a_{k, n}\bm{f} + b_{k, n}\mathbf{1},
\end{equation}
where $\bm{\varphi}^{\rm tr}_{k, n}$ is the approximated phase trend, $\bm{f}$ is the normalized subcarrier-frequency vector, and $a_{k, n}$ and $b_{k, n}$ denote the slope and intercept, respectively. The parameters $a_{k, n}$ and $b_{k, n}$ can be obtained by solving a weighted least-squares problem, which is given by
\begin{equation}
\min_{a_{k, n}, b_{k, n}}
\sum_{s=1}^{S} \zeta_{s, k, n}\left(\varphi_{s,k,n} - a_{k,n} f_s - b_{k,n}\right)^2,
\end{equation}
where the weight $\zeta_{s,k,n}$ is determined by the corresponding amplitude in the input channel $\widehat{\bm{H}}$ and given by
\begin{equation}
    \zeta_{s, k, n} = \max((\mu_{s, k, n})^p, \zeta_{\rm min}). 
\end{equation}
Here, $\mu_{s, k, n}$ is the amplitude of the $s, k, n$-th element in $\widehat{\bm{H}}$, $p$ is a predefined exponent, and $\zeta_{\rm min}$ is a small constant for numerical stability. In this way, unreliable low-amplitude phase samples contribute less to the trend estimation. As a result, the phase is decomposed into a linear trend component $\bm{\varphi}^{\rm tr}$ and a residual component $\bm{\varphi}^{\rm res}$ as
\begin{equation}
\bm{\varphi}^{\rm tr}_{k, n} = \bm{\varphi}_{k, n} - \bm{\varphi}^{\rm res}_{k, n}.
\end{equation}
Such a decomposition separates the channel phase into low-frequency (global) and high-frequency (local) components, which helps the subsequent transformer to better exploit both global consistency and fine-grained structural variations.

Based on the above decomposition, we construct two structural descriptors for token generation. The first is the trend descriptor $\bm{u}^{\rm tr}$ containing the average slope, mean channel amplitude, standard deviation of channel amplitude, and the variance of the trend component $\bm{\varphi}^{\rm tr}$. The second descriptor $\bm{u}^{\rm res}$, referred to as the residual descriptor, is composed of the energy of residual phase component $\bm{\varphi}^{\rm res}$, the variance of channel amplitude across subcarriers, mean channel amplitude, and the variance of the residual component $\bm{\varphi}^{\rm res}$. These two descriptors are then mapped into two learnable structural tokens as
\begin{align}
\bm{t}^{\rm tr} =~& \bm{e}^{\rm tr} + \mathcal{F}_{\rm tr}(\bm{u}^{\rm tr}) \in \mathbb{R}^{N_e},\\
\bm{t}^{\rm res} =~& \bm{e}^{\rm res} + \mathcal{F}_{\rm res}(\bm{u}^{\rm res}) \in \mathbb{R}^{N_e},
\end{align}
where $\bm{e}^{\rm tr}$ and $\bm{e}^{\rm res}$ are learnable token embeddings, and $\mathcal{F}_{\rm tr}(\cdot)$ and $\mathcal{F}_{\rm res}(\cdot)$ are single-layer MLPs for token generation.

\subsubsection{Latent Token Generation}

To generate latent tokens, the amplitude and phase of the input channel $\widehat{\bm{H}}$ are first computed, denoted by $\widehat{\bm{H}}^{\mathrm A}$ and $\widehat{\bm{H}}^{\mathrm P}$, respectively. The amplitude and phase tensors are then flattened into one-dimensional vectors over all subcarriers, receive antennas, and RIS elements, yielding
\begin{equation}
\widehat{\bm{h}}^{\rm AF}, \widehat{\bm{h}}^{\rm PF}\in\mathbb{R}^{1 \times (SKN)}.
\end{equation}
These two sequences are processed by two independent convolutional encoders with identical architecture. Specifically, each encoder consists of three $1$D convolutional layers with kernel sizes $5$, $3$, and $1$, respectively. Each convolutional layer is followed by a nonlinear activation and channel-wise layer normalization, while the extracted features are further compressed by max pooling and adaptive average pooling to extract compact feature representations. Finally, dropout and flattening are applied to obtain compact real-valued feature vectors. Denoting the amplitude and phase encoders by $\mathcal{F}_{\rm A}(\cdot)$ and $\mathcal{F}_{\rm P}(\cdot)$, respectively, we obtain
\begin{align}
\bm{f}_{\rm A} = \mathcal{F}_{\rm A}\left(\widehat{\bm{h}}^{\rm AF}\right),\\
\bm{f}_{\rm P} = \mathcal{F}_{\rm P}\left(\widehat{\bm{h}}^{\rm PF}\right),
\end{align}
where both $\bm{f}_{\rm A}$ and $\bm{f}_{\rm P}$ are the amplitude and phase feature vectors. They are concatenated and fused through a two-layer feed-forward network, and then projected into a sequence of latent tokens using a linear projection and reshaping. A learnable positional embedding $\bm P$ is also added to preserve token order, and the whole process is given by
\begin{equation}
\bm{T} = \mathcal{F}_{\rm T}\big([\bm{f}_{\rm A};\bm{f}_{\rm P}]\big) + \bm{P} \in \mathbb{R}^{N_t \times N_e},
\end{equation}
where $\bm{T} = (\bm{t}_1, \cdots, \bm{t}_{N_t})$ is the latent tokens, $\mathcal{F}_{\rm T}$ represents the tokenization process, $N_t$ is the token number, and $N_e$ is the embedding dimension. The final token sequence is thus formed as
\begin{equation}
\bm{T}^{\rm F} = \left(\bm{t}^{\rm tr}, \bm{t}^{\rm res},\bm{t}_1,\ldots,\bm{t}_{N_t}\right).
\end{equation}

\subsection{Transformer-Based Inference Module}
\label{ss_sim}

The inference module aims to convert the input tokens $\bm{T}^{\rm F}$ into the final semantic estimates, including the class label $\hat{c}$, the target center $\hat{\bm{r}}$, and the target size $\hat{\bm{s}}$. The key idea is to employ a transformer encoder to model the dependencies among tokens, while exploiting their distinct roles for classification and regression.

Specifically, the token sequence $\bm{T}^{\rm F}$ is fed into a multi-layer transformer encoder, where each layer consists of multi-head self-attention and feedforward sublayers. Denoting the transformer by $\mathcal{F}_{\rm trans}(\cdot)$, the output can be written as
\begin{equation}
\bm{Q} = \mathcal{F}_{\rm trans}\left(\bm{T}^{\rm F}\right),
\end{equation}
where $\bm{Q} = (\bm{q}^{\rm tr}, \bm{q}^{\rm res}, \bm{q}_1, \dots, \bm{q}_{N_t})$ preserves the same structure as the input sequence, where the first two output tokens correspond to the updated trend token $\bm{q}^{\rm tr}$ and residual token $\bm{q}^{\rm res}$, while others are latent tokens. Rather than directly using all latent tokens for inference, we extract three aggregated features from the latent tokens to capture target-related information. The first is the token-wise mean feature, given by
\begin{equation}
\bm{q}_{\rm mean} = \frac{1}{N_t}\sum^{N_t}_{i=1} \bm{q}_i,
\end{equation}
which captures the global average representation. The second is the token-wise max feature $\bm{q}_{\rm max}$, which emphasizes the most salient responses across tokens. Its $d$-th element is given by
\begin{equation}
[\bm{q}^{\rm max}]_d = \max_{i=1,\ldots,N_t} [\bm{q}_i]_d
\end{equation}
where $[\cdot]_d$ represents the $d$-th element of a vector. The third feature is the first latent token $\bm{q}_1$, which serves as an implicit summary of the token sequence through self-attention. These aggregated features can reduce redundancy among tokens while preserving both global context and salient local information for semantic sensing.

For target classification, the feature vector is constructed by concatenating the mean feature, the max feature, the first latent token, and the residual structural token, which is given by
\begin{equation}
\bm{q}^{\rm cls} =
\left(\bm{q}^{\rm mean}, \bm{q}^{\rm max}, \bm{q}_1, \bm{q}^{\rm res}\right).
\end{equation}
This vector is first processed by a feature fusion network consisting of a linear layer, layer normalization, and a nonlinear activation. The resulting features are then fed into a classifier composed of two fully connected layers, layer normalization, and ReLU activation, with a softmax output layer. The predicted class probabilities are given by
\begin{equation}
\widehat{\bm{p}} = \mathcal{F}_{\rm cls}\left(\bm{q}^{\rm cls}\right),
\end{equation}
where $\widehat{\bm{p}}$ denotes the predicted class probability vector. The use of the residual token in the classification head is motivated by the fact that target category is largely determined by fine-grained scattering structures and residual spectral variations rather than by coarse phase trend. Finally, the predicted class label is obtained by selecting the class with the highest predicted probability, given by
\begin{equation}
    \widehat{c} = \arg \max_{c} \widehat{p}_c.
\end{equation}

For target center localization, the regression feature is constructed from the trend token and the global mean latent feature, which is written as
\begin{equation}
\bm{q}^{\rm reg} =
\left(\bm{q}^{\rm tr}, \bm{q}^{\rm mean}\right).
\end{equation}
The estimated center location is obtained as
\begin{equation}
\widehat{\bm{r}} = \mathcal{F}_{\rm ctr}\left(\bm{q}^{\rm reg}\right),
\end{equation}
where $\mathcal{F}_{\rm ctr}$ is a lightweight regressor consisting of two linear layers, layer normalization, and ReLU activation.

The size estimation also starts from the regression feature $\bm{q}^{\rm reg}$, but it further incorporates the residual token through a gating mechanism. Specifically, a gate vector is first generated, given by
\begin{equation}
\bm{g} = \sigma(\bm{J} \bm{q}^{\rm res} + \bm{j}),
\end{equation}
where $\sigma(\cdot)$ denotes the sigmoid function, and $\bm{J}$ and $\bm{j}$ are learnable parameters. The gated regression feature is then expressed as
\begin{equation}
\bm{q}^{\rm size}
= \bm{q}^{\rm reg} + \eta(2\bm{g}-1),
\end{equation}
where $\eta$ is a scaling coefficient controlling the influence of the residual token. Finally, the target size is predicted by
\begin{equation}
\widehat{\bm{s}} = \mathcal{F}_{\rm size}\left(\bm{q}^{\rm size}\right),
\end{equation}
where $\mathcal{F}_{\rm size}$ is implemented as a two-layer fully connected network with layer normalization and nonlinear activation, and the last layer adopts a Softplus activation to ensure non-negative outputs. This residual-gated design is motivated by the different scale sensitivities of inference tasks. Target classification is primarily determined by fine-scale structural variations, whereas localization is governed by large-scale spatial characteristics. In contrast, target size is inherently a mid-scale attribute that depends on both global extent and local details. Hence, both the trend token and the residual token are jointly exploited for accurate size estimation.

\section{Joint Beamforming Design and Model Fine-Tuning}
\label{s_mstf}

In this section, we first provide the definition of semantic sensing loss to characterize the semantic sensing performance. Then, the corresponding optimization problem is formulated for joint beamforming design and model fine-tuning, where we aim to improve the semantic sensing performance under the communication rate constraint, followed by a multi-stage optimization method to solve the formulated problem.

\subsection{Semantic Sensing Loss}

Semantic sensing aims to jointly infer multiple target attributes from wireless signals. Therefore, the semantic sensing loss is defined as a weighted combination of classification, localization, and size estimation losses, i.e.,
\begin{equation}
\mathcal{L}
= \mathcal{L}_{\rm cls} +
\lambda_{\rm loc}\mathcal{L}_{\rm loc}
+
\lambda_{\rm size}\mathcal{L}_{\rm size},
\label{eq:total_loss}
\end{equation}
where $\mathcal{L}_{\rm cls}$, $\mathcal{L}_{\rm loc}$, and $\mathcal{L}_{\rm size}$ denote the classification, localization, and size estimation losses. The weighting coefficients $\lambda_{\rm loc}$ and $\lambda_{\rm size}$ are also introduced to balance among different losses. These losses are mathematically defined as
\begin{align}
\mathcal{L}_{\rm cls} &= -
\sum_{i=1}^{|\mathcal{C}|}
\nu_i \log \widehat{p}_i,
\label{eq:l_cls}\\
\mathcal{L}_{\rm loc} &= \big\| \widehat{\bm{r}}-\bm{r}^\star \big\|_2^2, \label{eq:l_loc} \\ \mathcal{L}_{\rm size} &= \big\| \widehat{\bm{s}}-\bm{s}^\star \big\|_2^2. \label{eq:l_size} \end{align} 
Here, the classification loss $\mathcal{L}_{\rm cls}$ in (\ref{eq:l_cls}) adopts the cross-entropy form, which is widely used to evaluate in classification tasks. In~(\ref{eq:l_cls}),  $|\mathcal{C}|$ denotes the number of semantic classes, $\nu_i \in \{0,1\}$ is the one-hot encoded ground-truth label, and $\widehat{p}_i$ is the predicted probability of the $i$-th class. The localization loss $\mathcal{L}_{\rm loc}$ and the size estimation loss $\mathcal{L}_{\rm size}$ are formulated using the squared $\ell_2$ norm.

\subsection{Problem Formulation}

The objective of the integrated semantic sensing and communication problem is to minimize the semantic sensing loss subject to communication rate and transmit power constraints. Alternative formulations are also possible. For example, the achievable communication rate can be maximized under sensing performance requirements. In this paper, to guarantee the communication rates of users, communication rate constraints are explicitly incorporated into the formulated problem while optimizing the semantic sensing performance, which is given by
\begin{subequations}
    \begin{align}
    {\rm P1:}&\min_{\bm{B}, \bm{\psi}, \bm{\theta}} \mathcal{L},\label{p1_obj}\\
    s.t.~& R_u \ge R^{\rm min}_u,\label{c_R_u}\\
    & \sum^S_{s=1} ||\bm{B}_{f, s}||^2 = P_0, \forall f,\label{c_P_0}\\
    & \psi_{f, n} \in [0, 2\pi),\label{c_psi}
\end{align}
\end{subequations}
where $\bm{B}$ and $\bm{\psi}$ are digital and RIS beamformers, respectively. Besides, $\bm{\theta} = \{\bm{\theta}_{\rm T}, \bm{\theta}_{\rm I}\}$ is the set of learnable parameters in the inference network, with $\bm{\theta}_{\rm T}$ and $\bm{\theta}_{\rm I}$ being the learnable parameters in the tokenization and inference modules, respectively. Parameter $R^{\rm min}_u$ is the communication rate threshold for the $u$-th user, $P_0$ is the transmit power of the BS, and $\psi_{f, n}$ are the phase shift of the $n$-th RIS element in the $f$-th frame. Constraint~(\ref{c_R_u}) ensures that the achievable communication rate of the $u$-th user is equal or greater than $R^{\rm min}_u$. Constraint~(\ref{c_P_0}) restricts the overall transmit power, and constraint~(\ref{c_psi}) sets the available value of RIS phase shift.

\subsection{Algorithm Design}

It is challenging to solve (P1) because both the objective function~(\ref{p1_obj}) and constraint~(\ref{c_R_u}) are highly non-linear. Moreover, the digital beamformer $\bm{B}$, RIS beamformer $\bm{\psi}$, and network parameters $\bm{\theta}$ are also coupled with each other. To effectively solve (P1), we first reformulate the problem to a simpler form, and then propose a multi-stage optimization algorithm to solve it efficiently.

Specifically, the communication rate constraint (\ref{c_R_u}) is first incorporated into the objective function through a penalty term, yielding the following optimization problem
\begin{subequations}
\begin{align}
{\rm P2:}&\min_{\bm{B}, \bm{\psi}, \bm{\theta}} \mathcal{L}_{\rm total} = \mathcal{L}
+
\lambda_{\rm dual} \cdot \frac{1}{U}\sum_{u=1}^{U} \left[R_u^{\min} - R_u\right]_+,
\label{eq:lagrangian_loss}\\
s.t.~& (\ref{c_P_0}), (\ref{c_psi}),\notag
\end{align}
\end{subequations}
where $[x]_+ \triangleq \max(x,0)$ denotes the hinge operator, and $\lambda_{\rm dual} \ge 0$ is a dual variable associated with the rate constraint. This formulation penalizes only the violation of the rate constraint, i.e., when $R_u < R_u^{\min}$, while no penalty is imposed if the constraint is satisfied. 

Next, to further reduce the optimization complexity of Problem (P1), we propose a multi-stage optimization algorithm to iteratively optimize the digital beamformer $\bm{B}$, the RIS beamformer $\bm{\Psi}$, and the network parameters $\bm{\theta}$. The key idea is to partially decouple the optimization of the inference network and the beamforming-related variables, such that different components can be optimized progressively before a final joint refinement. Specifically, the proposed method consists of three stages:
\begin{itemize}
    \item \textbf{Offline Pre-Training:} In the first stage, we optimize the learnable parameters $\bm{\theta}$ in the network. The digital beamformer $\bm{B}$ and RIS beamformer $\bm{\Psi}$ are not involved in this stage. Instead, the ground-truth channel data is input into the network, which is directly trained to learn the mapping from channel measurements to target attributes. The optimization problem is thus formulated as
    \begin{equation}
    \min_{\bm{\theta}}
    \ \mathcal{L}.
    \label{eq:stage1}
    \end{equation}
    Note that this stage is performed offline and aims to learn robust semantic feature representations independent of channel estimation, thereby providing a suitable initialization for subsequent stages.
    \item \textbf{Hybrid Beamforming Optimization:} In the second stage, we jointly optimize the digital beamformer $\bm{B}$ and RIS beamformer $\bm{\psi}$. Instead of directly optimizing the semantic sensing loss, we adopt a channel-level objective that facilitates stable training. Specifically, the loss function is defined as
\begin{align}
\mathcal{L}_{\rm ch}\! =~&\! \left|\left|\widehat{\bm{H}} \!-\! \bm{H}^\star\right|\right|_F^2 \!+\! \lambda_{\rm phy} \sum_{f, s}\left|\left|\bm{A}_{f, s} \left(\widehat{\bm{H}}_s\right)^{\mathsf H} \!-\! \bm{Z}^{\mathsf H}_{f, s}\right|\right|^2_F \notag\\
&+ \lambda_{\rm dual} \cdot \frac{1}{U}\sum_{u=1}^{U} \left[R_u^{\min} - R_u\right]_+,
\label{eq:channel_loss}
\end{align}
where $\widehat{\bm{H}}$ and $\bm{H}^\star$ denote the estimated and ground-truth channels, respectively, and $\widehat{\bm{H}}_s$ is the estimated channel at the $s$-subcarrier. The first term in (\ref{eq:channel_loss}) provides direct supervision on the channel estimation, while the second term enforces consistency with the underlying signal model, thereby improving robustness under limited or noisy observations. Finally, the third term is the penalty term to implement the communication rate constraint. Thus, the corresponding optimization problem is formulated as
\begin{align}
&\min_{\bm{B},\bm{\psi}}
\ \mathcal{L}_{\rm ch},\\
s.t.~&(\ref{c_P_0}), (\ref{c_psi}).\notag
\label{eq:stage2}
\end{align}
This problem can be efficiently solved using gradient descent method. To handle the transmit power constraint (\ref{c_P_0}), the beamforming matrix is normalized during the forward pass, give by
\begin{equation}
\bm{B}_{f, s} \leftarrow \sqrt{P_0} \frac{\bm{B}_{f, s}}{||\bm{B}_{f, s}||_F},
\end{equation}
In this way, the overall transmit power always satisfies the power budget. For constraint (\ref{c_psi}), the RIS phase variables are wrapped into the interval $[0,2\pi)$ by taking their modulo-
$2\pi$ values. Moreover, the dual variable $\lambda$ is updated iteratively following a primal-dual scheme. In particular, after each training iteration, $\lambda$ is updated as
\begin{equation}
\lambda_{\rm dual} \leftarrow \Big[
\lambda_{\rm dual} + \alpha_1 \cdot
\frac{1}{U}\sum_{u=1}^{U}
\big(
R_u^{\min} - R_u
\big)_+
\Big],
\label{eq:dual_update}
\end{equation}
where $\alpha_1 > 0$ is the dual step size. This update increases $\lambda_{\rm dual}$ when the rate constraint is violated and decreases its influence otherwise, thereby adaptively balancing channel estimation performance and communication requirements.
    \item \textbf{Joint Fine-Tuning:} In the first stage, the inference network is pre-trained using ground-truth channel information. However, target attribute inference relies on estimated channels obtained from the received signals under the optimized digital and RIS beamformers, and channel estimation errors may degrade the sensing performance. Therefore, joint fine-tuning is adopted to improve the end-to-end inference performance against channel estimation errors, which is formulated as
    \begin{align}
    &\min_{\bm{B},\bm{\psi},\bm{\theta}}
    \ \mathcal{L}_{\rm total},\\
    s.t.& (\ref{c_P_0}), (\ref{c_psi}).
    \label{eq:stage3}
    \end{align}
    Here, the constraints (\ref{c_P_0}) and (\ref{c_psi}) are enforced in the same manner as in the previous stage. The pre-trained network parameters, together with the optimized digital and RIS beamformers obtained in the previous stages, are used as initialization.
\end{itemize}

\section{Simulation Results}
\label{s_nr}

\subsection{Simulation Setup}

\begin{table}[t]
\centering
\caption{Main Simulation Parameters}
\renewcommand{\arraystretch}{1.4}
\label{tab_simulation_parameters}
\begin{tabular}{ll}
\hline
\textbf{Parameter} & \textbf{Value} \\
\hline
BS antenna array size & $8\times1$ \\
RIS size & $16\times16$ \\
PR antenna array size & $2\times1$ \\
Carrier frequency & $3.5$ GHz \\
Maximum BS transmit power & $30$ dBm \\
Noise power & $-104$ dBm \\
Communication rate threshold $R_u^{\rm min}$ & $5$ bit/s/Hz \\
Voxel size & $0.1\times0.1\times0.1$ m$^3$ \\
Maximum target size & $1.6\times1.6\times1.6$ m$^3$ \\
Learning rate & $5\times10^{-4}$ \\
Batch size & $32$ \\
Number of latent/trend/residual tokens & $32/1/1$ \\
Token embedding dimension & $128$ \\
Number of Transformer layers & $4$ \\
Transformer feedforward dimension & $512$ \\
\hline
\end{tabular}
\vspace{-4mm}
\end{table}

In the considered wireless system, the BS employs an $8\times1$ antenna array, the RIS contains $16\times16$ metamaterial elements, while the receiver is equipped with a $2\times1$ antenna array. The system operates at a carrier frequency of $3.5$ GHz. 
The maximum transmit power at the BS is $30$ dBm, and the noise powers at both the receiver and the communication user are set to $-104$ dBm~\cite{wang2025ris}. The communication rate threshold $R^{\rm min}_u$ is set to $5$ bit/s.

The dataset for integrated semantic sensing and communication is constructed using targets from ShapeNet~\cite{chang2015shapenet}, containing six target categories, i.e., car, chair, lamp, plane, sofa, and table. Each target is discretized into voxels with the size of each voxel being $0.1\times 0.1\times 0.1$ m$^3$. The maximum size of an target is $1.6\times1.6\times1.6$ m$^3$. The sensing region ranges from $2$ m to $7$ m along the $x$-axis and from $-2.5$ m to $2.5$ m along the $y$-axis. This dataset is randomly divided into training, validation, and test sets with a ratio of $80\%$, $10\%$, and $10\%$, respectively. The models are trained using the Adam optimizer with a learning rate of $5\times10^{-4}$ and a batch size of $32$. To mitigate the impact of class imbalance, a weighted random sampling strategy is adopted during training, where the sampling probability of each sample is inversely proportional to the number of samples in its corresponding class. 

In the inference network, the input signals and RIS beamformer are transformed into $32$ latent tokens, $1$ trend token and $1$ residual token, with embedding dimension of each token being $128$. The Transformer encoder consists of $4$ layers  with feedforward dimension $512$ and dropout probability $0.1$. The feature fusion network in the classification head contains a linear projection layer with dimensions $512\rightarrow128$, and the dimensions of the classifier are $128\rightarrow64\rightarrow6$. As for the localization head, the dimensions of the linear projection layer in the feature fusion network are $256\rightarrow128$, and the dimensions of the regressor are $128\rightarrow128\rightarrow2$. In size estimation head, the dimensions of $\mathcal{F}_{\rm size}$ is $128\rightarrow64\rightarrow3$.

The performance is evaluated in terms of classification accuracy, localization error, size estimation error, and semantic sensing loss. The localization error and size estimation error are measured using the root mean square error (RMSE) between the estimated and ground-truth values. Main simulation parameters are summarized in Table~\ref{tab_simulation_parameters}.

\subsection{Performance Comparison}

\begin{table*}[t]
\centering
\caption{Semantic Sensing Performance Comparison Under Same Communication Rate Threshold}
\renewcommand{\arraystretch}{1.4}
\label{tab_compare}
\begin{tabular}{l|ccc|ccc}
\hline
\multirow{2}{*}{Method}

& \multicolumn{3}{c|}{$\bm{F = 1}$} & \multicolumn{3}{c}{$\bm{F = 2}$}\\

\cline{2-7}

& \textbf{Acc.} & \textbf{Loc. Err. [m]} & \textbf{Size Err. [m]} & \textbf{Acc.} & \textbf{Loc. Err. [m]} & \textbf{Size Err. [m]}\\
\hline
CS & $-$ & 3.8361 & 1.4672 & $-$ & 2.7578 & 1.2152 \\
CS + CNN & 0.4461 & 0.8103 & 0.5755 & 0.5928 & 0.5427 & 0.5587 \\
Dual-branch & 0.7743 & 0.4805 & $-$ & 0.7886 & 0.4738 & $-$ \\
\hline
\textbf{Proposed} & \textbf{0.8625} & \textbf{0.3352} & \textbf{0.2643} & \textbf{0.9121} & \textbf{0.2354} & \textbf{0.2273}\\
\hline
\end{tabular}
\vspace{-4mm}
\end{table*}

In this subsection, we compare the proposed scheme with three benchmark methods under the same communication rate constraint. Specifically, the following benchmark schemes are considered.
\begin{itemize}
    \item \textbf{Compressed sensing~(CS):} In this scheme, a CS-based sparse reconstruction algorithm~\cite{eldar2012compressed} is employed to reconstruct the scattering vector $\bm{v}$ given the transmitted and received signals. Due to the existence of noise and reconstruction errors in the recovered $\bm{v}$, a threshold is introduced to identify effective voxels belonging to the target. The target center and size are then estimated from the selected voxels. Moreover, the threshold is optimized using the training dataset to minimize the localization and size estimation errors.
    \item \textbf{CS + CNN:} This benchmark first reconstructs the scattering vector using the same CS-based method. Then, a CNN-based inference network is adopted to estimate the target class, location, and size from the reconstructed scattering vector. This network consists of two 3D convolutional layers with ReLU activation and max-pooling operations, followed by a shared fully connected feature extraction layer. Three task-specific heads are then employed for different tasks.
    \item \textbf{Dual-branch:} This method~\cite{zhao2026extended} employs two task-specific branches to infer class and location attributes from wireless channels, respectively. Specifically, a semantic branch is adopted for target classification, while a geometric branch is used for localization. Cross-task attention is further introduced to exchange complementary features between the two branches. Target size estimation is not considered in this work.
\end{itemize}

Table~\ref{tab_compare} compares the semantic sensing performance of different methods under different numbers of OFDM frames. It can be observed that the proposed method consistently achieves the best performance in terms of classification accuracy, localization error, and size estimation error. For example, when $F=2$, the accuracies of CS+CNN and dual-branch are only $59.28\%$ and $78.86\%$, respectively, while the proposed scheme achieves a classification accuracy of $91.21\%$, demonstrating that accurate semantic sensing can be achieved with very limited sensing measurements. In addition, the localization error of the proposed method is reduced to $0.2354$m, which is significantly lower than those of CS and CS+CNN. Overall, compared with existing methods, the proposed scheme achieves more than $12\%$ performance gains for all three types of object-level information under the same communication rate constraint when $F = 2$.

Moreover, increasing the number of sensing frames from $F=1$ to $F=2$ improves the performance of all schemes. This is because different beamformers across multiple frames provide additional observation diversity. Nevertheless, the CS-based methods still exhibit significantly larger localization and size estimation errors compared with the proposed scheme. For example, even when $F=2$, the localization and size estimation errors of CS remain substantially higher than those of the proposed method. This is because the neighboring voxels produce highly correlated channel responses, making it difficult for the CS algorithm to accurately distinguish the scattering components of different voxels. As a result, the recovered voxels may contain noticeable reconstruction errors, which further degrade the localization and size estimation performance. In addition, as the CS+CNN scheme also relies on the reconstructed voxels as the input of the CNN, the reconstruction errors introduced by the CS further propagate to the inference stage, thereby resulting in a performance gap compared with the proposed end-to-end semantic sensing framework.

\begin{figure}[!t]
  \centering
  \includegraphics[width=2.8in]{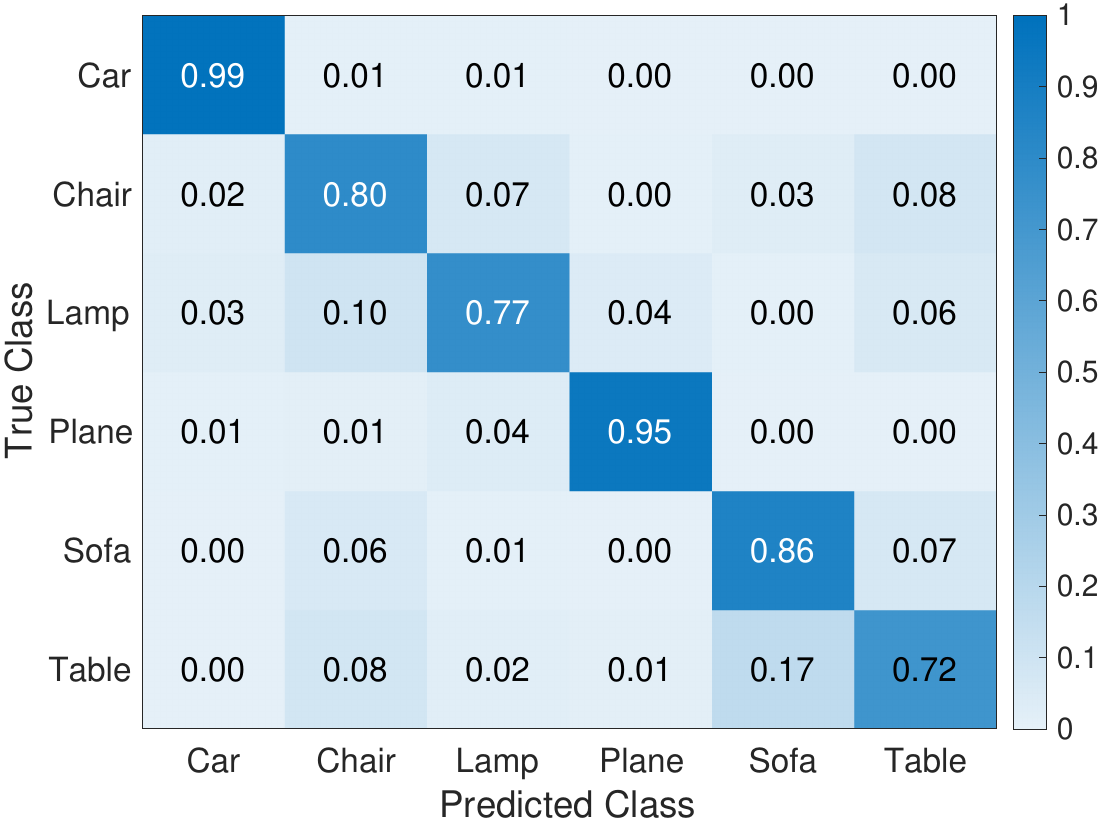}
  \vspace{-2mm}
  \caption{The confusion matrix obtained by the proposed method given $F = 1$.}
  \vspace{-2mm}
  \label{f_confusion_f1}
\end{figure}

\begin{figure}[!t]
  \centering
  \includegraphics[width=2.8in]{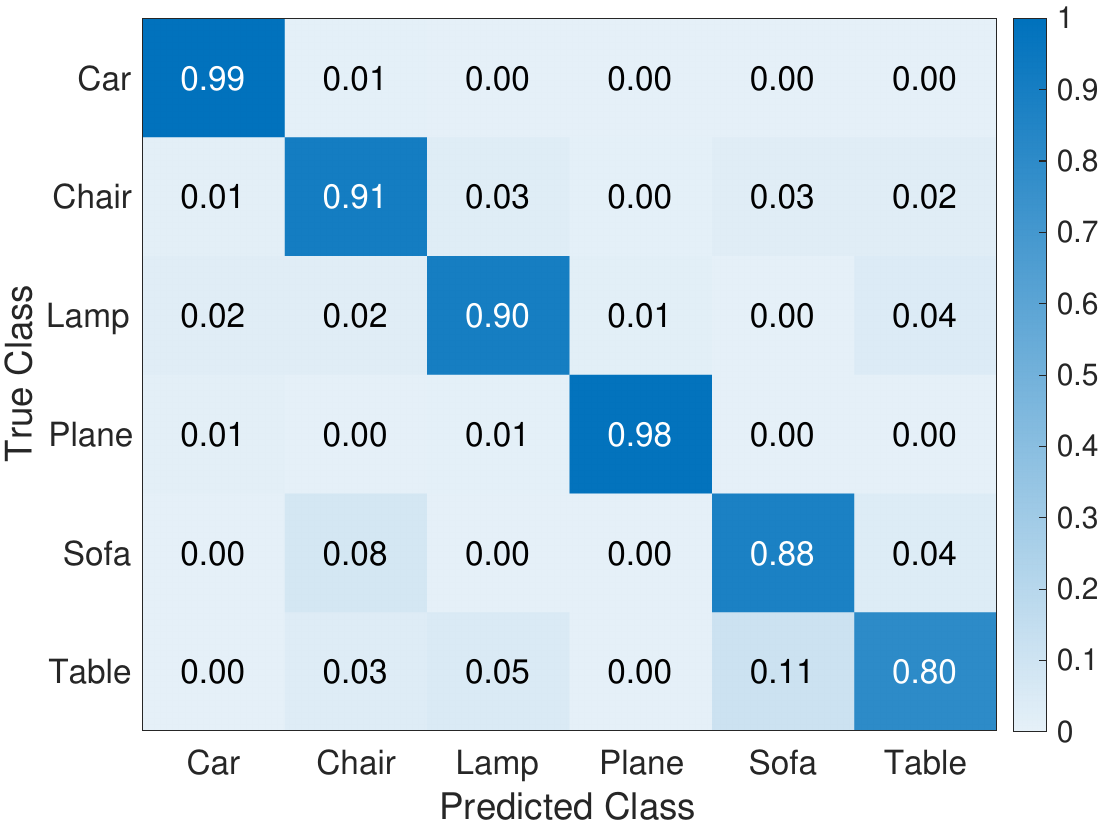}
  \vspace{-2mm}
  \caption{The confusion matrix obtained by the proposed method given $F = 2$.}
  \vspace{-2mm}
  \label{f_confusion_f2}
\end{figure}

Figs.~\ref{f_confusion_f1} and \ref{f_confusion_f2} illustrate the confusion matrices of the proposed method under $F=1$ and $F=2$, respectively. It can be observed that most target categories can be accurately recognized under both settings, where the classification accuracies of car and plane remain particularly high. Moreover, increasing the number of sensing frames from $F=1$ to $F=2$ further improves the classification performance for almost all target categories. For example, the classification accuracies of chair and lamp increase from $80\%$ and $77\%$ to $91\%$ and $90\%$, respectively. In contrast, sofa and table exhibit lower accuracies compared with other categories due to their more similar geometric structures. Nevertheless, the classification accuracies of all target categories remain at least 80\% when $F=2$, which demonstrates the effectiveness of the proposed method for multi-class classification.

\begin{figure}[!t]
  \centering
  \includegraphics[width=2.8in]{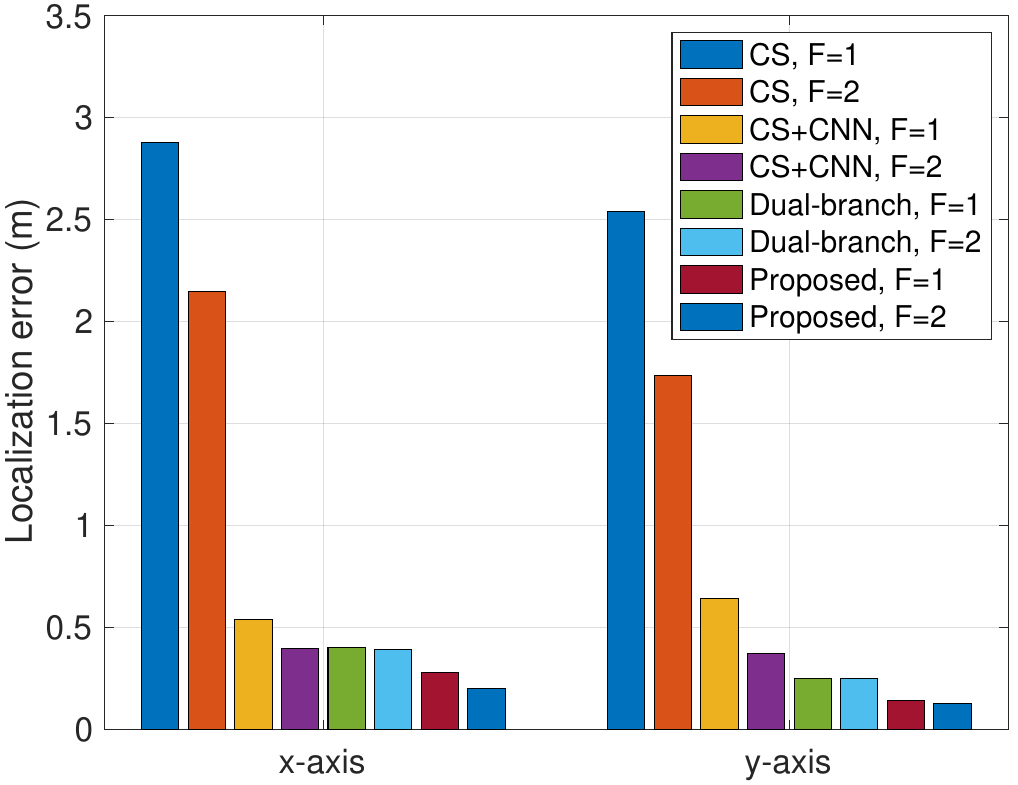}
  \vspace{-2mm}
  \caption{The localization error along x- and y-axes.}
  \vspace{-2mm}
  \label{f_compare_loc}
\end{figure}

Fig.~\ref{f_compare_loc} further compares the localization errors along the x- and y-axes, and we assume all targets rest on a horizontal ground plane
defined by $z=0$. It can be observed that the proposed method achieves the lowest localization errors on both axes. In addition, the localization error along the x-axis is generally larger than that along the y-axis for all schemes, and increasing the number of sensing frames from $F=1$ to $F=2$ reduces the localization errors of all methods on both axes.

\begin{figure}[!t]
  \centering
  \includegraphics[width=2.8in]{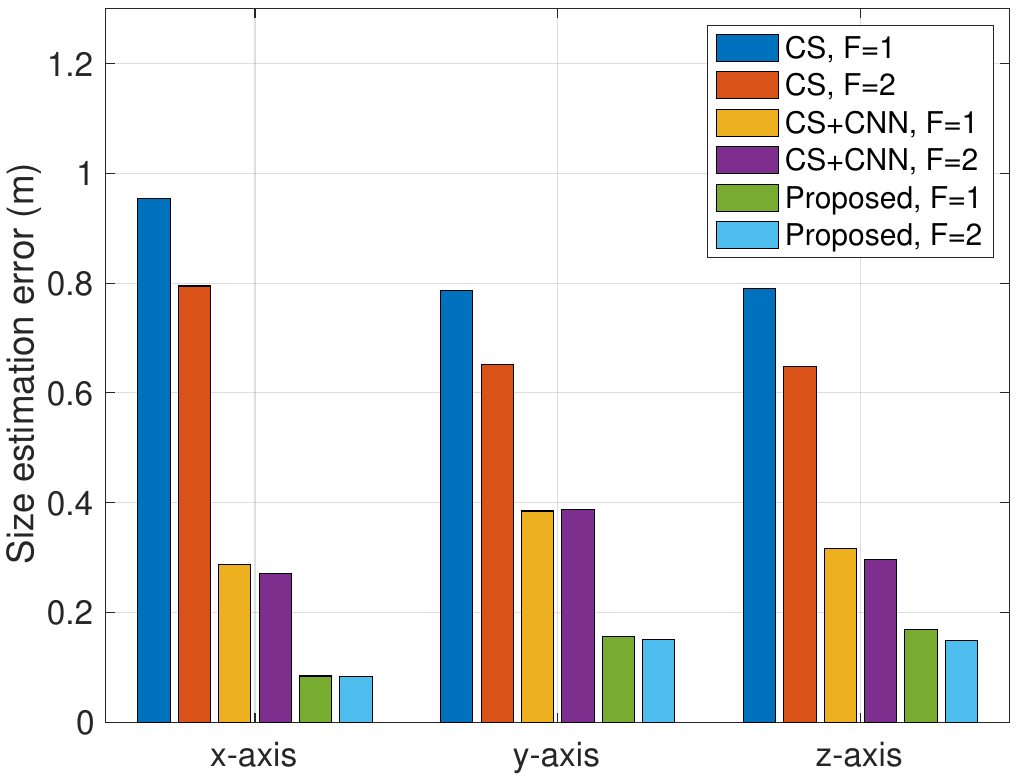}
  \vspace{-2mm}
  \caption{The size estimation error along x-, y-, and z-axes.}
  \vspace{-2mm}
  \label{f_compare_size}
\end{figure}

The size estimation performance along different axes is shown in Fig.~\ref{f_compare_size}. Similar to the localization results, the proposed method achieves the lowest estimation errors along all three axes. Moreover, compared with the localization errors in Fig.~\ref{f_compare_loc}, the size estimation errors exhibit smaller performance improvement when the number of frames increases from $F=1$ to $F=2$. This indicates that the localization performance is more sensitive to the increase of frames under the considered setup. 

\begin{figure*}[!t]
\centering
\subfloat[]{\includegraphics[width=2.2in]{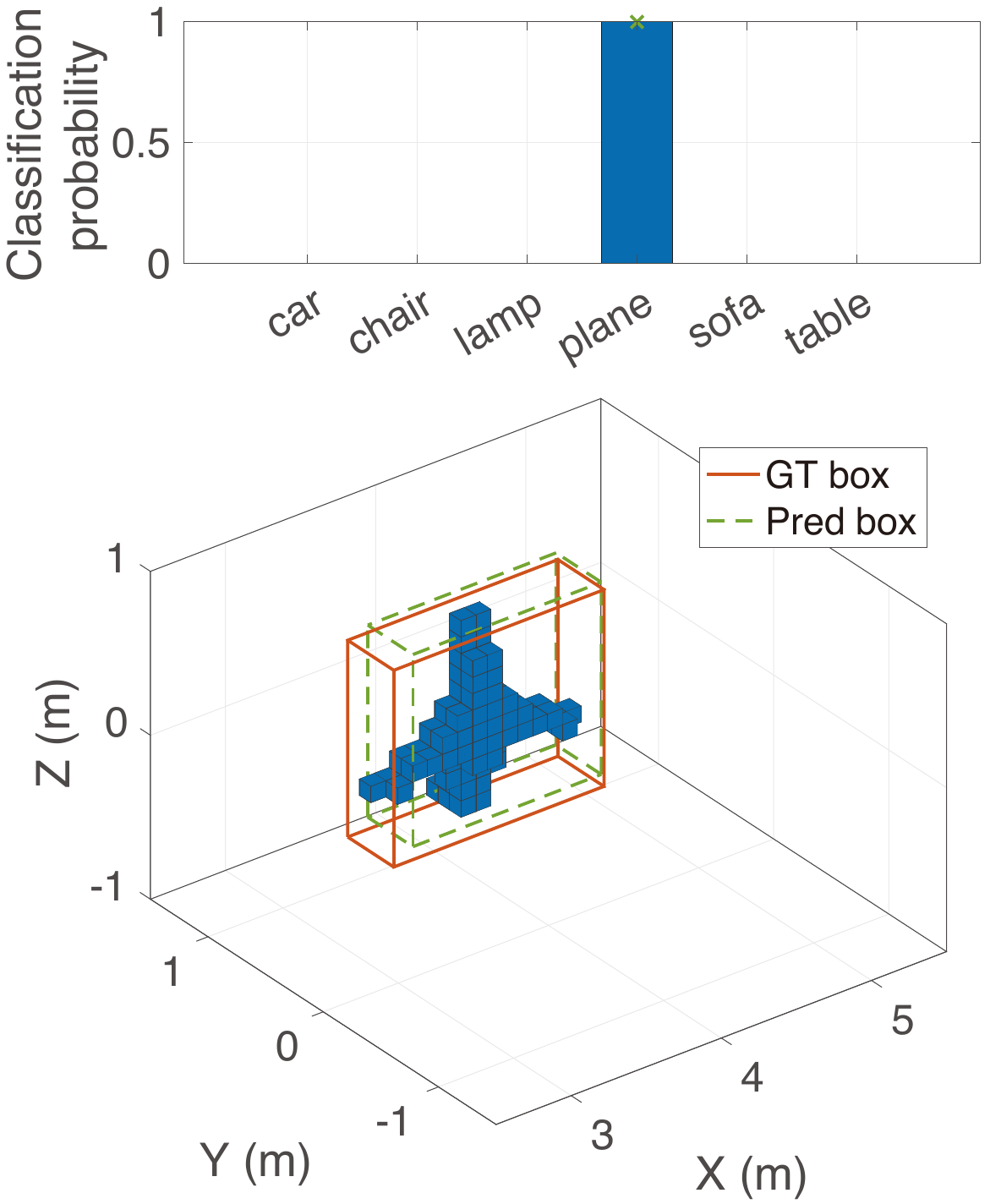}%
}
\hspace{4mm}
\subfloat[]{\includegraphics[width=2.2in]{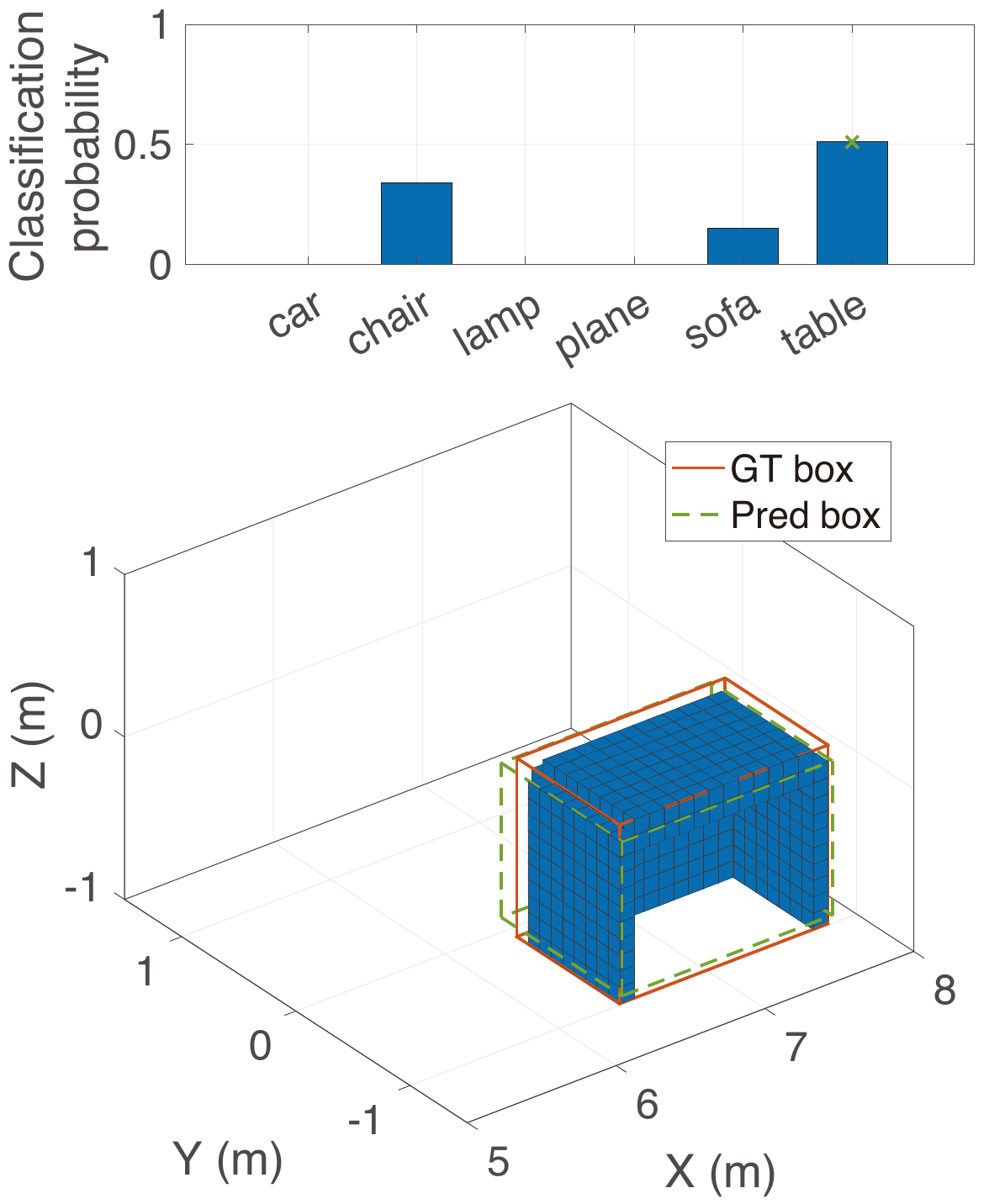}%
}
\hspace{4mm}
\subfloat[]{\includegraphics[width=2.2in]{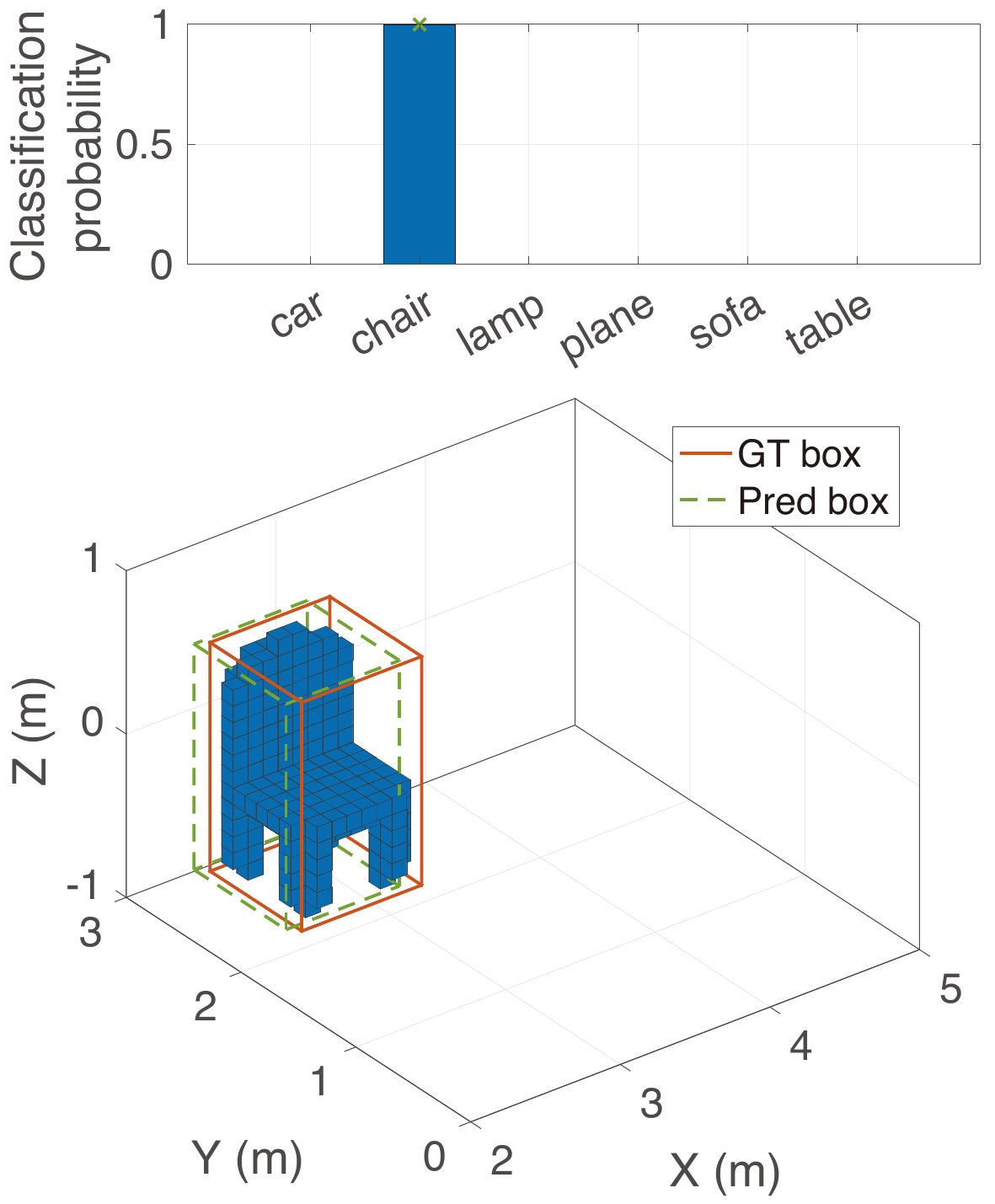}%
}
\vspace{-1mm}
\caption{Illustration of the classification, localization, and size estimation results: (a) A plane; (b) A table; (c) A chair.}
\vspace{-4mm}
\label{f_sample}
\end{figure*}

Semantic sensing results for three representative samples are illustrated in Fig.~\ref{f_sample}. Here, the predicted probability for each class indicates the model's confidence that the sensed target belongs to that class. In addition, the red and green bounding boxes represent the ground-truth~(GT) and predicted bounding boxes calculated from the corresponding target locations and sizes, respectively. It can be observed that the correct target category consistently achieves the highest classification probability for all samples. Meanwhile, the predicted bounding boxes closely match the ground-truth ones in both spatial position and geometric size. These results demonstrate the effectiveness of accurate joint classification, localization, and size estimation without explicit reconstruction of target shapes.

\subsection{Communication-Semantic Sensing Tradeoff}

\begin{figure}[!t]
  \centering
  \includegraphics[width=2.6in]{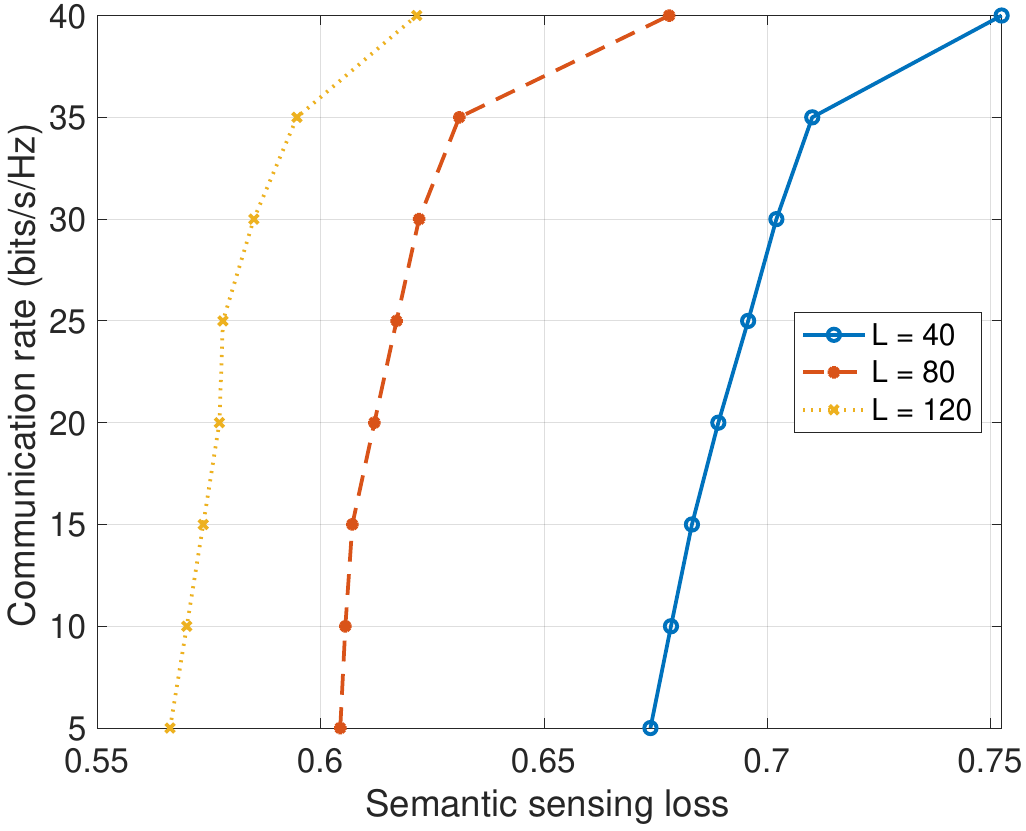}
  \vspace{-2mm}
  \caption{The communication rate per user versus the semantic sensing loss given different numbers of symbols.}
  \vspace{-2mm}
  \label{f_rate_loss}
\end{figure}

Fig.~\ref{f_rate_loss} illustrates the tradeoff between semantic sensing accuracy and achievable communication rate under different numbers of OFDM symbols when $U = 2$ and $P_0 = 10$dBm. It can be observed that the achievable communication rate per user generally increases with the semantic sensing loss. This is because improving communication performance requires the transmitted beams to be more focused on enhancing the communication links, thereby reducing the semantic sensing capability. In addition, increasing the number of OFDM symbols shifts the tradeoff curve toward the left. This is because a larger number of OFDM symbols provides more sensing observations, allowing a given semantic sensing accuracy to be maintained with beamformers that better support communication.

\begin{figure}[!t]
  \centering
  \includegraphics[width=2.6in]{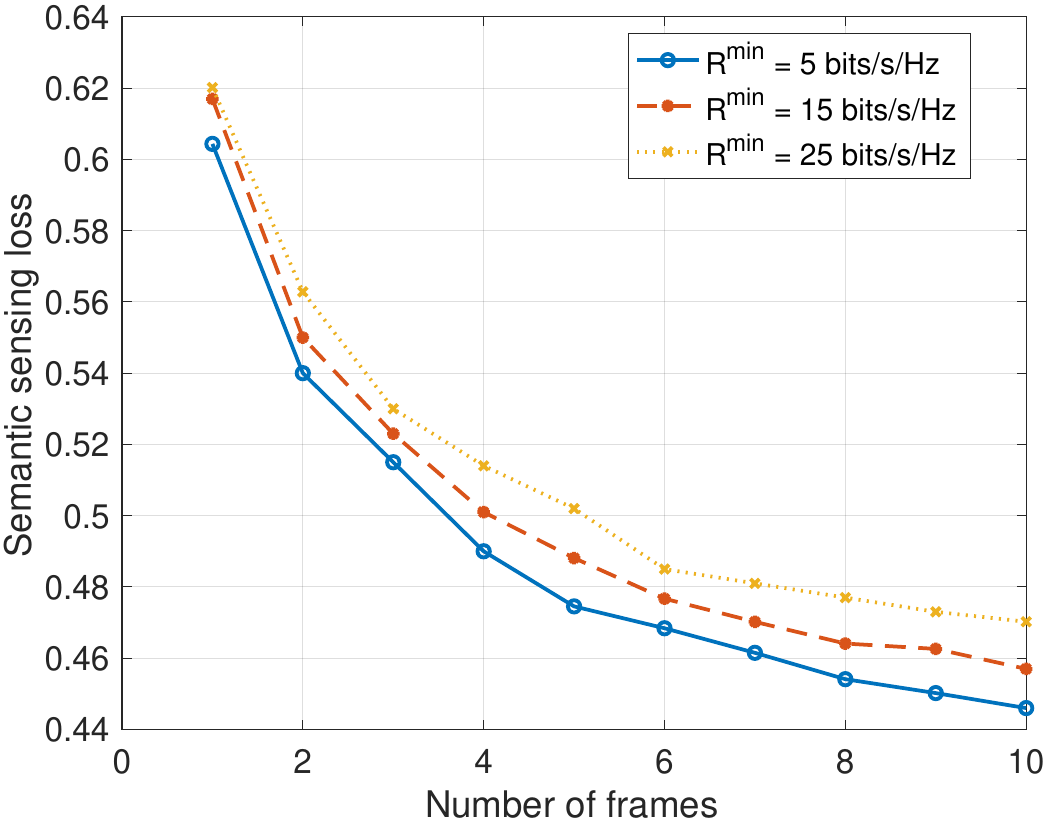}
  \vspace{-2mm}
  \caption{The semantic sensing loss versus the number of frames given different communication rate threshold.}
  \vspace{-4mm}
  \label{f_loss_F}
\end{figure}

Fig.~\ref{f_loss_F} shows the semantic sensing loss versus the number of sensing frames under different communication rate thresholds when $U = 2$ and $P_0 = 10$dBm. It can first be observed that the semantic sensing loss increases with the communication rate threshold. Moreover, for all considered communication rate thresholds, the semantic sensing loss decreases as the number of sensing frames increases, while the performance improvement gradually saturates when the number of frames is sufficiently large. This shows that most of the performance gains are achieved with only a few sensing frames, suggesting that accurate semantic sensing can be realized with low measurement overhead.

\subsection{Ablation Study}

Table~\ref{tab_ablation} presents the ablation study of the proposed scheme, where the impacts of the hybrid beamforming optimization module, the trend-residual guided tokenization module, and the Transformer-based inference module are respectively evaluated. Specifically, the first variant replaces the optimized digital and RIS beamformers with a random digital beamformer and RIS beamformers, while retaining the proposed tokenization and inference modules. The second variant replaces the proposed tokenization module with a CNN-based tokenizer. The third variant replaces the Transformer-based inference module and the associated prediction heads with a MLP-based predictor. The last row corresponds to the original proposed scheme.

It can be observed that the proposed scheme achieves the best performance across all three metrics, which verifies the effectiveness of jointly designing the beamforming optimization, tokenization, and Transformer-based inference modules. Among the three modules, the beamforming optimization module has the most significant impact on the overall performance. Specifically, when the optimized digital and RIS beamformers are replaced with random ones, the localization error increases from $0.2354$ m to $0.7054$ m, while the classification accuracy decreases from $91.21\%$ to $77.46\%$. Moreover, replacing the proposed tokenization module also leads to noticeable degradation in all three metrics. In contrast, removing the Transformer-based inference module mainly degrades the localization performance, while its impact on the classification accuracy and size estimation error is smaller. This observation is consistent with the characteristics of Transformer architectures, since the self-attention mechanism is particularly effective in capturing global dependencies among latent tokens. Such global feature modeling is especially important for localization tasks, where the target position is jointly determined by the overall spatial characteristics of the sensed scene rather than local features.

\setcellgapes{3pt}
\makegapedcells
\begin{table*}[t]
\centering
\caption{Ablation study of the proposed scheme}
\renewcommand{\arraystretch}{1.4}
\label{tab_ablation}
\begin{tabular}{ccc|cccc}
\hline
\makecell{\textbf{Beamforming} \\ \textbf{optimization}}
& \textbf{Tokenization} & \textbf{Inference} & \textbf{Loss} & \textbf{Acc.} & \textbf{Loc. Err. [m]} & \textbf{Size Err. [m]}\\
\hline
$\times$ & \checkmark & \checkmark & 0.7742 & 0.7746 & 0.7054 & 0.3475 \\
\checkmark & $\times$ & \checkmark & 0.5358 & 0.8331 & 0.3962 & 0.2875 \\
\checkmark & \checkmark & $\times$ & 0.3953 & 0.8820 & 0.3253 & 0.2732 \\
\hline
\checkmark & \checkmark & \checkmark & \textbf{0.2899} &\textbf{0.9121} & \textbf{0.2354} & \textbf{0.2273} \\
\hline
\end{tabular}
\vspace{-3mm}
\end{table*}

\vspace{-2mm}
\section{Conclusion}
\label{s_c}

In this paper, we have investigated an integrated semantic sensing and communication system, where a RIS-aided BS simultaneously performs semantic sensing and multi-user communication. Different from conventional ISAC frameworks that mainly focus on estimating specific sensing parameters, the proposed framework directly infers multiple target attributes, including target class, location, and size, from the received ISAC signals. To achieve this objective, a physics-guided Transformer-based target attribute inference network has been proposed, where trend and residual tokens were generated to jointly capture large-scale propagation characteristics and fine-grained scattering structures. In addition, a multi-stage training framework was developed to jointly design the digital and RIS beamformers and fine-tune the inference model under multi-user communication constraints. Simulation results have shown that: 1) the proposed method achieves better performance in terms of classification accuracy, localization error, and size estimation error under the same communication rate constraint. More specifically, it can realize 91.21\% classification accuracy using only two OFDM frames, demonstrating the feasibility of accurate semantic sensing with low measurement overhead; 2) there exists an inherent communication-semantic sensing tradeoff in the ISAC systems, while increasing the number of sensing frames enables a more favorable tradeoff by reducing the semantic sensing loss under different communication rate thresholds; 3) different semantic sensing tasks exhibit different sensitivities to the additional observation diversity introduced by multiple frames. In particular, increasing the number of frames provides larger improvement for localization compared with classification and size estimation.

\vspace{-1mm}
\bibliographystyle{IEEEtran}
\bibliography{myReference.bib}

@ARTICLE{zhang2025target,
  author={Zhang, Zhenkun and Ren, Hong and Pan, Cunhua and Hong, Sheng and Wang, Dongming and Wang, Jiangzhou and You, Xiaohu},
  journal={IEEE Trans. Commun.}, 
  title={Target Localization in Cooperative {ISAC} Systems: A Scheme Based on {5G} {NR} {OFDM} Signals}, 
  year={2025},
  volume={73},
  number={5},
  pages={3562-3578},
  month=may
}

@ARTICLE{guo2025multiges,
  author={Guo, Zhengxin and Wang, Dongzi and Cui, Kaiyan and Gui, Linqing and Ye, Ning and Xiao, Fu},
  journal={IEEE J. Sel. Areas Commun.}, 
  title={{MultiGes}: Real-Time Multi-Target Gesture Recognition for {ISAC}-Driven Human–Computer Interaction}, 
  year={2026},
  volume={44},
  number={},
  pages={899-912},
  month=sep}

@ARTICLE{zhang2026feature,
  author={Zhang, Haobo and Di, Boya and Zhang, Hongliang and Niyato, Dusit and Song, Lingyang},
  journal={IEEE Trans. Wirel. Commun.}, 
  title={Feature-Foundation Model Evolution for Low-Latency Semantic Communication}, 
  year={2026},
  volume={25},
  number={},
  pages={11915-11931},
  month=feb}

@ARTICLE{hu2024cross,
  author={Hu, Jingzhi and Niyato, Dusit and Luo, Jun},
  journal={IEEE J. Sel. Areas Commun.}, 
  title={Cross-Domain Learning Framework for Tracking Users in {RIS}-Aided Multi-Band ISAC Systems With Sparse Labeled Data}, 
  year={2024},
  volume={42},
  number={10},
  pages={2754-2768},
  month=oct
}

@ARTICLE{he2026task,
  author={He, Yinghui and Li, Xin and Luo, Jun},
  journal={IEEE Trans. Mobile Comput.}, 
  title={Task-Oriented Integrated Sensing and Semantic Communications for Multi-Device Video Analytics}, 
  year={2026},
  volume={25},
  number={5},
  pages={7323-7337},
  month=may}

@ARTICLE{zhang2021metalocalization,
  author={Zhang, Haobo and Zhang, Hongliang and Di, Boya and Bian, Kaigui and Han, Zhu and Song, Lingyang},
  journal={IEEE Trans. Wirel. Commun.}, 
  title={MetaLocalization: Reconfigurable Intelligent Surface Aided Multi-User Wireless Indoor Localization}, 
  year={2021},
  volume={20},
  number={12},
  pages={7743-7757},
  month           = {Dec.}
}

@ARTICLE{zeng2024ris,
  author={Zeng, Shuhao and Di, Boya and Zhang, Hongliang and Gao, Jiahao and Yue, Shaohua and Hu, Xinyuan and Fu, Rui and Zhou, Jiaqi and Liu, Xu and Zhang, Haobo and Wang, Yuhan and Sun, Shaohui and Qin, Haichao and Su, Xin and Wang, Mengjun and Song, Lingyang},
  journal={IEEE Wirel. Commun.}, 
  title={{RIS}-Based {IMT}-2030 Testbed for MmWave Multi-Stream Ultra-Massive {MIMO} Communications}, 
  year={2024},
  volume={31},
  number={3},
  pages={375-382},
  month=jun
}

@ARTICLE{zhang2026holographic,
  author={Zhang, Haobo and Zeng, Shuhao and Hu, Xinyuan and Zhang, Shupei and Di, Boya and Zhang, Hongliang and Li, Lianlin and Chang, Lin and Jiang, Wei Xiang and Sun, Shaohui and Han, Zhu and Eldar, Yonina C. and Zhang, Rui and Debbah, Mérouane and Song, Lingyang},
  journal={IEEE Trans. Commun.}, 
  title={Holographic Surface Enabled Integrated Sensing and Communications}, 
  year={2026},
  volume={},
  number={},
  month={early access}
}

@ARTICLE{sheng2025single,
  author={Sheng, Xianjun and Xu, Lining and Liu, Ning and Li, Hui},
  journal={IEEE Antennas Wirel. Propag. Lett.}, 
  title={Single-Layer Dual-Resonance Wideband Electrical Reconfigurable Intelligent Surface Element Design Using Characteristic Mode Analysis}, 
  year={2025},
  volume={24},
  number={11},
  pages={4298-4302},
  month=nov}

@article{hu2022metasketch,
  title={MetaSketch: Wireless semantic segmentation by reconfigurable intelligent surfaces},
  author={Hu, Jingzhi and Zhang, Hongliang and Bian, Kaigui and Han, Zhu and Poor, H Vincent and Song, Lingyang},
  journal={IEEE Trans. Wirel. Commun.},
  volume={21},
  number={8},
  pages={5916--5929},
  year={2022},
  month=aug
}

@ARTICLE{fredj2024channel,
  author={Fredj, Firas and Feriani, Amal and Mezghani, Amine and Hossain, Ekram},
  journal={IEEE Trans. Wirel. Commun.}, 
  title={Channel Estimation in {RIS}-Enabled mmWave Wireless Systems: A Variational Inference Approach}, 
  year={2024},
  volume={23},
  number={8},
  pages={10350-10365},
  month=aug}

@ARTICLE{yang2025integrated,
  author={Yang, Yinchao and Yang, Zhaohui and Huang, Chongwen and Xu, Wei and Zhang, Zhaoyang and Niyato, Dusit and Shikh-Bahaei, Mohammad},
  journal={IEEE Trans. Veh. Technol.}, 
  title={Integrated Sensing, Computing and Semantic Communication for Vehicular Networks}, 
  year={2025},
  volume={74},
  number={11},
  pages={18307-18312},
  month=nov}

@ARTICLE{bazzi2025low,
  author={Bazzi, Ahmad and Chafii, Marwa},
  journal={IEEE J. Sel. Areas Commun.}, 
  title={Low Dynamic Range for {RIS}-Aided Bistatic Integrated Sensing and Communication}, 
  year={2025},
  volume={43},
  number={3},
  pages={912-927},
  month=mar
}

@article{zhao2026extended,
  title={Extended-Target Classification and Localization for Near-Field {ISAC}},
  author={Zhao, Zongyao and Wang, Zhaolin and Han, Lincong and Jin, Jing and Liu, Yuanwei and Huang, Kaibin},
  journal={arXiv preprint arXiv:2603.23093},
  year={2026}
}

@INPROCEEDINGS{rosinol2020kimera,
  author={Rosinol, Antoni and Abate, Marcus and Chang, Yun and Carlone, Luca},
  booktitle={Proc. IEEE Int. Conf. Robot. Autom. (ICRA)}, 
  title={Kimera: an Open-Source Library for Real-Time Metric-Semantic Localization and Mapping}, 
  year={2020},
  month=may,
  address={Paris, France}}

@INPROCEEDINGS{shakoor2025simultaneous,
  author={Shakoor, Usama and Janjua, Muhammad Bilal and Solaija, Muhammad Sohaib J. and Arslan, Hüseyin},
  booktitle={Int. Balkan Conf. Commun. Netw. (Balkancom)}, 
  title={Simultaneous Intrusion Detection and Localization Using {ISAC} Network}, 
  year={2025},
  address={Piraeus, Greece},
  month=jun}

@INPROCEEDINGS{tosi2024feasibility,
  author={Tosi, Paolo and Henninger, Marcus and de Oliveira, Lucas Giroto and Mandelli, Silvio},
  booktitle={nt. Workshop Signal Process. Advances Wirel. Commun. (SPAWC)}, 
  title={Feasibility of Non-Line-of-Sight Integrated Sensing and Communication at mmWave}, 
  year={2024},
  month=sep,
  address={Lucca, Italy}}

@ARTICLE{zhang2026integrated,
  author={Zhang, Di and Cui, Yuanhao and Cao, Xiaowen and Su, Nanchi and Gong, Yi and Liu, Fan and Yuan, Weijie and Jing, Xiaojun and Andrew Zhang, J. and Xu, Jie and Masouros, Christos and Niyato, Dusit and Renzo, Marco Di},
  journal={IEEE Commun. Surv. Tutor.}, 
  title={Integrated Sensing and Communications Over the Years: An Evolution Perspective}, 
  year={2026},
  month={early access}
}

@ARTICLE{zeng2026holograhpic,
  author={Zeng, Shuhao and Zhang, Haobo and Di, Boya and Zhang, Hongliang and Shao, Zijian and Han, Zhu and Vincent Poor, H. and Song, Lingyang},
  journal={IEEE J. Sel. Areas Commun.}, 
  title={Holographic Beamforming for Integrated Sensing and Communication With Mutual Coupling Effects}, 
  year={2026},
  volume={44},
  number={},
  pages={480-497},
  month=sep}

@ARTICLE{zhong2023joint,
  author={Zhong, Kai and Hu, Jinfeng and Pan, Cunhua and Deng, Minglong and Fang, Jun},
  journal={IEEE Signal Process. Lett.}, 
  title={Joint Waveform and Beamforming Design for {RIS}-Aided {ISAC} Systems}, 
  year={2023},
  volume={30},
  number={},
  pages={165-169},
  month=feb}

@ARTICLE{zhang2025joint,
  author={Zhang, Guoxin and Wang, Cong and Yi, Wei and Ngo, Hien Quoc and Matthaiou, Michail},
  journal={IEEE Trans. Aerosp. Electron. Syst.}, 
  title={Joint Detection and Localization of Moving Targets in Multi-Cell {ISAC} Systems}, 
  year={2025},
  volume={},
  number={},
  pages={1-20},
  month={early access}}

@INPROCEEDINGS{wu2023machine,
  author={Wu, Tong and Wang, Cheng-Xiang and Li, Junling and Huang, Chen},
  booktitle={Proc. IEEE Int. Symp. Pers., Indoor Mobile Radio Commun. (PIMRC)}, 
  title={Machine Learning-based Predictive Channel Modeling for {6G} Wireless Communications Using Image Semantic Segmentation}, 
  year={2023},
  address={Toronto, Canada},
  month=sep}

@ARTICLE{zhang2022holographic,
  author={Zhang, Haobo and Zhang, Hongliang and Di, Boya and Renzo, Marco Di and Han, Zhu and Poor, H. Vincent and Song, Lingyang},
  journal={IEEE J. Sel. Areas Commun.}, 
  title={Holographic Integrated Sensing and Communication}, 
  year={2022},
  volume={40},
  number={7},
  pages={2114-2130},
  month=jul
}

@INPROCEEDINGS{gao2026generative,
  author={Gao, Jiahao and Zhang, Haobo and Di, Boya and Song, Lingyang},
  booktitle={Proc. Int. Conf. Commun. (ICC)}, 
  title={Generative {AI}-driven Wireless Semantic Sensing by the Dual-polarized Reconfigurable Intelligent Surface}, 
  year={2026},
  address={Glasgow, UK},
  month={May}
}

@book{eldar2012compressed,
  title={Compressed sensing: Theory and applications},
  author={Eldar, Yonina C and Kutyniok, Gitta},
  year={2012},
  address={Cambridge, UK},
  publisher={Cambridge University Press}
}

@INPROCEEDINGS{zhang2022parameter,
  author={Zhang, Xiaoyu and Zhang, Haobo and Zhang, Hongliang and Liu, Liang and Di, Boya},
  booktitle={2022 Int. Symp. Wireless Commun. Syst. (ISWCS)}, 
  title={Parameter Estimation for Reconfigurable Holographic Surfaces enabled Radars}, 
  year={2022},
  month={Oct.},
  address         = {Hangzhou, China},
}

@ARTICLE{zhu2024cramer,
  author={Zhu, Qi and Li, Ming and Liu, Rang and Liu, Qian},
  journal={IEEE Trans. Wirel. Commun.}, 
  title={Cramér-{Rao} Bound Optimization for Active {RIS}-Empowered {ISAC} Systems}, 
  year={2024},
  volume={23},
  number={9},
  pages={11723-11736},
  month=sep}

@ARTICLE{xiao2026channel,
  author={Xiao, Jian and Wang, Ji and Cui, Qimei and Yang, Yucang and Li, Xingwang and Niyato, Dusit and Yuen, Chau},
  journal={IEEE Trans. Wirel. Commun.}, 
  title={Channel Estimation for Flexible Intelligent Metasurfaces: From Model-Based Approaches to Neural Operators}, 
  year={2026},
  volume={25},
  number={},
  pages={10684-10701},
  month=jan}

@ARTICLE{xiao2025sparsity,
  author={Xiao, Zichao and Liu, Rang and Li, Ming and Wang, Wei and Liu, Qian},
  journal={IEEE Trans. Commun.}, 
  title={Sparsity Exploitation via Joint Receive Processing and Transmit Beamforming Design for {MIMO-OFDM} {ISAC} Systems}, 
  year={2025},
  volume={73},
  number={5},
  pages={3593-3607},
  month={May}}

@ARTICLE{yin2025deep,
  author={Yin, Mingxi and Huang, Min and Hu, Rui and Xu, Hao},
  journal={IEEE Wirel. Commun. Lett.}, 
  title={Deep Learning-Assisted {ISAC} for {UAV} Detection in Multipath Environments With Micro-Doppler}, 
  year={2025},
  volume={14},
  number={9},
  pages={2862-2866},
  month=sep}

@ARTICLE{zhang2022metaradar,
  author={Zhang, Haobo and Zhang, Hongliang and Di, Boya and Bian, Kaigui and Han, Zhu and Song, Lingyang},
  journal={IEEE Trans. Wirel. Commun.}, 
  title={{MetaRadar}: Multi-Target Detection for Reconfigurable Intelligent Surface Aided Radar Systems}, 
  year={2022},
  volume={21},
  number={9},
  pages={6994-7010},
  month=sep}

@article{chang2015shapenet,
  title={{ShapeNet}: An information-rich {3D} model repository},
  author={Chang, Angel X and Funkhouser, Thomas and Guibas, Leonidas and Hanrahan, Pat and Huang, Qixing and Li, Zimo and Savarese, Silvio and Savva, Manolis and Song, Shuran and Su, Hao and others},
  journal={arXiv preprint arXiv:1512.03012},
  year={2015}
}

@ARTICLE{egilmez2017graph,
  author={Egilmez, Hilmi E. and Pavez, Eduardo and Ortega, Antonio},
  journal={IEEE J. Sel. Topics Signal Process.}, 
  title={Graph Learning From Data Under Laplacian and Structural Constraints}, 
  year={2017},
  volume={11},
  number={6},
  pages={825-841},
  month=sep}

@INPROCEEDINGS{wang2025ris,
  author={Wang, Yuhan and Zhang, Haobo and Liu, Qingyu and Zhang, Hongliang and Song, Lingyang},
  booktitle={Proc. Asia-Pacific Conf. Commun. (APCC)}, 
  title={{RIS}-Aided Wireless Amodal Sensing for Single-View {3D} Reconstruction}, 
  year={2025},
  month=nov,
  address={Osaka, Japan}
}

\vfill

\end{document}